\documentclass[trackchanges,twocolumn, twocolappendix]{aastex701}
\usepackage{CJK}
\usepackage{float}
\usepackage{multirow}

\begin{document}
\begin{CJK*}{UTF8}{gbsn}

\title{Unveiling the Sources of X-ray Luminosity in DESI Galaxy Groups: Insights from the SRG/eROSITA All-Sky Survey}

\correspondingauthor{Yun-Liang Zheng}

\author[0000-0002-5632-9345]{Yun-Liang Zheng (郑云亮)}
\affiliation{Department of Astronomy, School of Physics and Astronomy, and Shanghai Key Laboratory for Particle Physics and Cosmology, Shanghai Jiao Tong University, Shanghai 200240, People's Republic of China}
\email[show]{zhengyunliang@sjtu.edu.cn}  

\author[0000-0003-3997-4606]{Xiaohu Yang (杨小虎)}
\affiliation{Department of Astronomy, School of Physics and Astronomy, and Shanghai Key Laboratory for Particle Physics and Cosmology, Shanghai Jiao Tong University, Shanghai 200240, People's Republic of China}
\affiliation{State Key Laboratory of Dark Matter Physics, Tsung-Dao Lee Institute, Shanghai Jiao Tong University, Shanghai 200240, People's Republic of China}
\email{} 

\author[0000-0002-2941-6734]{Teng Liu (刘腾)}
\affiliation{Department of Astronomy, University of Science and Technology of China, Hefei 230026, People’s Republic of China}
\affiliation{School of Astronomy and Space Science, University of Science and Technology of China, Hefei 230026, People’s Republic of China}
\email{}

\author[0009-0004-9950-9807]{Shijiang Chen (陈世江)}
\affiliation{Department of Astronomy, University of Science and Technology of China, Hefei 230026, People’s Republic of China}
\affiliation{School of Astronomy and Space Science, University of Science and Technology of China, Hefei 230026, People’s Republic of China}
\email{}

\author[]{Esra Bulbul}
\affiliation{Max-Planck-Institut f\"ur extraterrestrische Physik (MPE), Gie\ss{}enbachstra\ss{}e 1, D-85748 Garching bei M\"unchen, Germany}
\email{}

\author[0000-0003-3501-0359]{Ang Liu (刘昂)}
\affiliation{Institute for Frontiers in Astronomy and Astrophysics, Beijing Normal University, Beijing 102206, People's Republic of China}
\affiliation{Max-Planck-Institut f\"ur extraterrestrische Physik (MPE), Gie\ss{}enbachstra\ss{}e 1, D-85748 Garching bei M\"unchen, Germany}
\email{}

\author[]{Yi Zhang (张艺)}
\affiliation{Max-Planck-Institut f\"ur extraterrestrische Physik (MPE), Gie\ss{}enbachstra\ss{}e 1, D-85748 Garching bei M\"unchen, Germany}
\email{}

\author[0009-0000-5534-1935]{Dawei Li (李大为)}
\affiliation{Department of Astronomy, Xiamen University, Xiamen, Fujian 361005, People's Republic of China}
\email{}

\author[0000-0002-5458-4254]{Xi Kang (康熙)}
\affiliation{Institute for Astronomy, School of Physics, Zhejiang University, Hangzhou 310027, People's Republic of China}
\affiliation{Center for Cosmology and Computational Astrophysics, Zhejiang University, Hangzhou 310027, People's Republic of China}
\affiliation{Purple Mountain Observatory, 10 Yuan Hua Road, Nanjing 210034, People's Republic of China}
\email{}

\author[0000-0003-3196-7938]{Yizhou Gu (顾一舟)}
\affiliation{State Key Laboratory of Dark Matter Physics, Tsung-Dao Lee Institute, Shanghai Jiao Tong University, Shanghai 200240, People's Republic of China}
\email{}

\author[0000-0003-3203-3299]{Yirong Wang (王艺蓉)}
\affiliation{Department of Astronomy, School of Physics and Astronomy, and Shanghai Key Laboratory for Particle Physics and Cosmology, Shanghai Jiao Tong University, Shanghai 200240, People's Republic of China}
\affiliation{State Key Laboratory of Dark Matter Physics, Tsung-Dao Lee Institute, Shanghai Jiao Tong University, Shanghai 200240, People's Republic of China}
\email{}

\author[0000-0003-0771-1350]{Qingyang Li (李清洋)}
\affiliation{Department of Astronomy, School of Physics and Astronomy, and Shanghai Key Laboratory for Particle Physics and Cosmology, Shanghai Jiao Tong University, Shanghai 200240, People's Republic of China}
\email{}

\author[]{Jiaqi Wang (王佳琪)}
\affiliation{Department of Astronomy, School of Physics and Astronomy, and Shanghai Key Laboratory for Particle Physics and Cosmology, Shanghai Jiao Tong University, Shanghai 200240, People's Republic of China}
\affiliation{Institute for Computational Cosmology, Department of Physics, Durham University, South Road, Durham, DH1 3LE, UK}
\email{}



\begin{abstract}
We use the first eROSITA all-sky survey (eRASS1) to investigate the contributions of AGN and extended gas to the total X-ray luminosity ($L_X$) of galaxy groups with different halo masses ($M_h$) at different redshifts. The presence of AGN in their central galaxies is identified using multi-wavelength catalogs, including the X-ray counterparts, the ASKAP radio catalog, and the DESI spectroscopic measurements. We apply the stacking method to obtain sufficient statistics for the X-ray surface brightness profile and the $L_X$ for groups with different central AGN properties. We find that the X-ray groups exhibit the highest $L_X$, followed by groups with QSO, radio, BPT-AGN, and non-AGN centrals. Moreover, the $L_X$ of the $M_h \lesssim 10^{13}h^{-1}M_\odot$ groups is dominated by the central AGN, while the X-ray emission from extended gas tends to be more prominent in the $M_h \gtrsim 10^{13}h^{-1}M_\odot$ groups. In groups where the AGN play a major role in X-ray emission, the contribution from extended gas is minor, resulting in significant uncertainties concerning the extended X-ray emission. When the subset containing the X-ray detected counterparts is excluded, the extended gas component becomes easier to obtain. A correlation has been identified between the X-ray luminosity of the central AGN and extended gas. However, once we account for the positional offset, their correlation becomes less prominent. Currently, the results are not conclusive enough to confirm whether there is a connection between the AGN feedback and extended gas. However, they provide a new perspective on the feedback processes in the history of group assembly.
\end{abstract}


\section{Introduction}

A galaxy group\footnote{In this paper, we refer to a system of galaxies as a group regardless of its mass and richness, whether it is a massive cluster containing many galaxies, a small group with only a few members, or even an isolated single galaxy. Within the framework of the halo model, all galaxies in such a system are assumed to reside in a dark matter halo.} is a concentration of galaxies assumed to be embedded within an extended dark matter halo, providing probes of the spatial distribution and growth history of large-scale structure \citep[e.g.][]{Yang..2005, Yang..2007, Yang..2012}. The relatively higher density of these regions provides an ideal site for studying large-scale structure formation and evolution from an observational perspective \citep[e.g.,][]{Weinmann..2006, Coenda..2012, Wetzel..2012, Knobel..2015, Davies..2019, Raouf..2019, Cluver..2020, Zheng..2022, Bahar..2024}. In addition to the stellar and dust components provided by the galaxy members, gas typically constitutes the majority of the baryonic matter in a group. In the context of gravitational theory, the distribution of temperature and mass of gas is anticipated to align self-similarly with the group's potential well. Nevertheless, baryonic feedbacks associated with galaxy evolution can significantly influence the gas component within the halo. Feedback from active supermassive black holes (SMBHs) is widely considered as a key mechanism in regulating star formation in central galaxies and increasing the temperature of the surrounding gas \citep[e.g.,][]{Fabian2012,Zinn..2013,Silk..2014}. Outflows and jets from the central active galactic nucleus (AGN) interact with the surrounding hot intragroup medium (IGrM), injecting substantial energy that inhibits efficient cooling of the gas and suppresses subsequent star formation.

According to the halo mass function, the mass density of the nearby universe is dominated by halos with masses around $M_h \sim 10^{13-13.5}h^{-1}M_\odot$ \citep[e.g.,][]{Driver..2022, CuiWG2024, Dev..2024}. While the conditional luminosity function model indicates that star formation efficiency peaks at $M_h \sim 10^{12}h^{-1}M_\odot$ \cite[e.g.,][]{Yang..2003}. Beyond this mass threshold, the effects of the AGN feedback gradually become significant \citep[e.g.,][]{Fanidakis..2013, Chauhan..2020, Byrne..2024, Eckert..2024}. Therefore, studying the central galaxies in halos within this mass range can help us understand the impact of baryonic feedback on galaxy evolution, including its impact to the surrounding halo gas.

Observationally, the X-ray emission from galaxy groups is roughly contributed by three components: diffuse hot gas, AGN, and X-ray binaries \citep{Vladutescu-Zopp..2023, Marini..2024, Vladutescu-Zopp..2025}. Massive groups are primarily dominated by extended hot gas, whereas the X-ray emission from smaller groups is predominantly contributed by AGN. Some massive groups/clusters ($M_h \gtrsim 5-14 \times 10^{13}h^{-1}M_\odot$) that can be detected directly in the X-ray map generally have a hot gas component that dominates overwhelmingly, and their X-ray brightness typically shows spherical symmetry. In contrast, the X-ray emission from most low-mass halos ($M_h \lesssim 10^{13}h^{-1}M_\odot$) is extremely weak, making it unlikely to be detected directly in X-ray surveys. As the results shown by \citet{Rasmussen..2006}, \citet{Anderson..2015}, \citet{Andreon..2016}, and \citet{O'Sullivan..2017}, X-ray surveys miss a large fraction of galaxy groups, indicating that X-ray–selected samples might not be fully representative of the underlying population. In a small fraction of groups with multiple galaxy members and no strongly dominant central galaxy, the X-ray image often exhibits multiple peaks. The alignment of these peaks with the positions of galaxy members suggests that they are most likely driven by AGN activity \citep[e.g.,][]{Mulchaey2000, Desjardins..2013, O'Sullivan..2014, Morandi..2017}.

Due to the faint X-ray emissions of low-mass groups, detecting their X-ray signals demands exceptionally high observational sensitivity. Although longer exposure times can enhance detection rates, X-ray telescopes still have notable limitations. X-ray telescopes such as XMM-Newton and Chandra offer deep exposures but are limited to relatively small sky coverage, rendering them inadequate for the detection of large survey samples \citep[][]{Brandt..2005}. Other X-ray telescopes, such as ROSAT, have conducted all-sky surveys but with relatively shallow exposures, typically averaging a few hundreds of seconds \citep[][]{Xue2017}. ROSAT also have a wide PSF, which make it hard to detect compact faint groups. As a result, the direct identification of low-mass groups that represent the dominant population of group systems through blind searches in all-sky X-ray images remains highly challenging, except in cases where individual sources exhibit exceptionally high luminosities.

Although observational evidence relate to small groups has gradually accumulated in recent years, the completeness of X-ray detections for low-mass groups remains limited \citep[e.g.,][]{Lovisari2021, Damsted..2024}. This limitation becomes even more apparent when using optically selected groups as priors for measuring X-ray luminosity, highlighting the difficulty of achieving completeness when identifying groups directly in X-rays \citep[e.g.,][]{Zheng..2023, Popesso..2024s, Marini..2025}. Although non-gravitational heating effects, such as AGN feedback, have a significant impact on the X-ray luminosity \citep{Anderson..2015}, it remains challenging at present to observationally disentangle the respective X-ray contributions from the AGN and the hot gas in low-mass groups. Some AGNs are very bright in X-ray and can be directly blind-detected in X-ray maps, but this method of identifying AGN is inherently highly biased \citep{Azadi..2017}. Generally, the most commonly used method to identify AGN are as follows: X-ray selection \citep[e.g.,][]{Ranalli..2005, Menzel..2016, Birchall..2023}, near- and mid-infrared color \citep[e.g.,][]{Donley..2012}, spectral energy distribution (SED) fitting \citep[e.g.,][]{Brammer..2008, Boquien..2019, YangG..2020}, spectroscopic classification \citep[e.g.,][]{Kewley..2001, Kauffmann..2003, Mignoli..2013}, and radio selection \citep[e.g.,][]{Magliocchetti..2018, Hardcastle..2025}. As pointed out by \citet{Mazzolari..2024}, the first three diagnostics are more often applied to the selection of radiatively efficient AGN. Nevertheless, once heavily obscured AGN are included, the efficiency of X-ray and infrared selection methods shows significant limitations. Therefore, disentangling the AGN contribution to the X-ray emission is far from straightforward, as properties such as AGN and hot gas components in these groups remain poorly constrained. This observational challenge has made the study of X-ray properties in low-mass groups heavily dependent on cosmological simulations, while simultaneously increasing the demand for more sensitive observational instruments.

Since 2019, the new generation all-sky X-ray telescope, eROSITA, was launched and began the entire sky scanning in the $0.2-10$ keV range across eight full-sky scans. This is achieved through an array of seven aligned telescope modules (TMs), with an average sky exposure time of $\sim 2500$ seconds. Prior to the commencement of the first survey, the eROSITA Final Equatorial-Depth Survey (eFEDS) was specifically designed to assess the capabilities of the eROSITA mission. The sky coverage of eFEDS is $\sim 140$ deg$^2$, and is overlapped with several deep optical/NIR surveys such as the HSC Wide Area Survey \citep{Aihara..2018}, KIDS-VIKINGS \citep{Kuijken..2019}, the DESI Legacy Imaging Survey \citep{Dey..2019}, etc. However, \citet{Brunner..2022} indicate that within this small area, $\sim$ 30,000 sources can be directly detected in X-ray maps, of which only 542 are clearly extended sources emitted by hot gas in massive galaxy clusters with redshifts up to $z \simeq 1.32$ \citep{LiuA..2022}. \citet{Bulbul..2022} highlighted that due to the eROSITA point-spread function (PSF), some dimmer galaxy clusters could be erroneously classified as point sources by the pipeline. They compiled an additional catalog containing 346 X-ray point sources, which are actually galaxy groups/clusters with masses ranging from $10^{13}$ to $4.5 \times 10^{14} M_{\odot}$. Nonetheless, the completeness of this sample for galaxy clusters within this mass range remains significantly less than that of group samples selected in the optical/NIR bands. Therefore, the most direct method of studying the X-ray properties of complete galaxy group samples is to combine optical/NIR galaxy surveys with all-sky X-ray surveys \citep[e.g.,][]{Bahar..2022, Zheng..2023,Popesso..2023,Popesso..2024,Zhang..2024a,Zhang..2024b,Zhang..2024c,LiDW..2024}. The former provides reliable galaxy group samples, which can be used as prior information to probe their X-ray properties.

Recently, \citet{Zheng..2023} utilized the DESI group catalog from \citet{Yang..2021} , which includes $\sim 600$k groups in the eFEDS region with a minimum halo mass reaching $M_h \simeq 10^{10.75} h^{-1}M_\odot$ and a maximum redshift of $z \simeq 1$, to perform X-ray measurements. Nearly half of the low-mass groups exhibit X-ray signals too faint to yield effective X-ray measurements. Therefore, to enhance the X-ray signal, we use stacking techniques to obtain the X-ray emissions for galaxy groups with different optical properties. We find that the average soft band $L_{\rm X}$ is correlated with $M_h$, but the slope is shallower than the self-similar prediction. Moreover, it was also found that the average $L_{\rm X}$ of low-mass groups is largely influenced by the AGNs in their central galaxies. Because of restricted sky coverage and the fact that not all DESI galaxy samples have been subjected to spectroscopic observations, the overall sample size remains inadequate. As a result, it is challenging to identify which galaxies have strong AGN components, and disentangling the different contributions to the X-ray emission becomes unfeasible.

The first eROSITA all-sky survey (eRASS1) has now been publicly released, covering the western Galactic hemisphere. Although the average exposure time across most of the footprint is only $\sim\!1/10$ of that in eFEDS, the sample of DESI groups within the eROSITA coverage has increased dramatically -- from $\sim$ 600k to more than 40 million. Utilizing spectroscopic data from 2.5 million central galaxies in DESI groups that overlap with eRASS1, along with emission lines derived from DESI optical spectra \citep{DESI..2024b}, blind X-ray detections provided by eRASS1 \citep{Merloni..2024}, and radio galaxy data from ASKAP \citep{Hale..2021}, we are now better equipped to thoroughly examine how AGN characteristics influence the X-ray emissions of galaxy groups across a range of halo masses based on larger group samples and more detailed multi-wavelength observations

The organization of this paper is as follows: Section~\ref{sec:catalogs} introduces the optically selected group samples used for measuring the X-ray emission, along with the multi-waveband observational data. In Section~\ref{sec:class}, we categorize different group types based on the central galaxy features pertinent to our research. Section~\ref{sec:results} details the mean X-ray surface brightness profiles and X-ray luminosities for various group components. In Section~\ref{sec:discuss}, we delve into how observational effects might influence our findings. The conclusion is provided in Section~\ref{sec:conclusion}. For clarity in conveying the paper's main points, the X-ray stacking method and the approach for distinguishing the contributions of different components to the X-ray emission are detailed in the Appendices.

Throughout this paper, we assume a flat $\Lambda$CDM cosmology with parameters: $\Omega_m = 0.3089$, $\Omega_b = 0.0486$, and $H_0 = 100h$ km/s/Mpc.

\begin{figure*}
    \centering
    \includegraphics[width=1.\hsize]{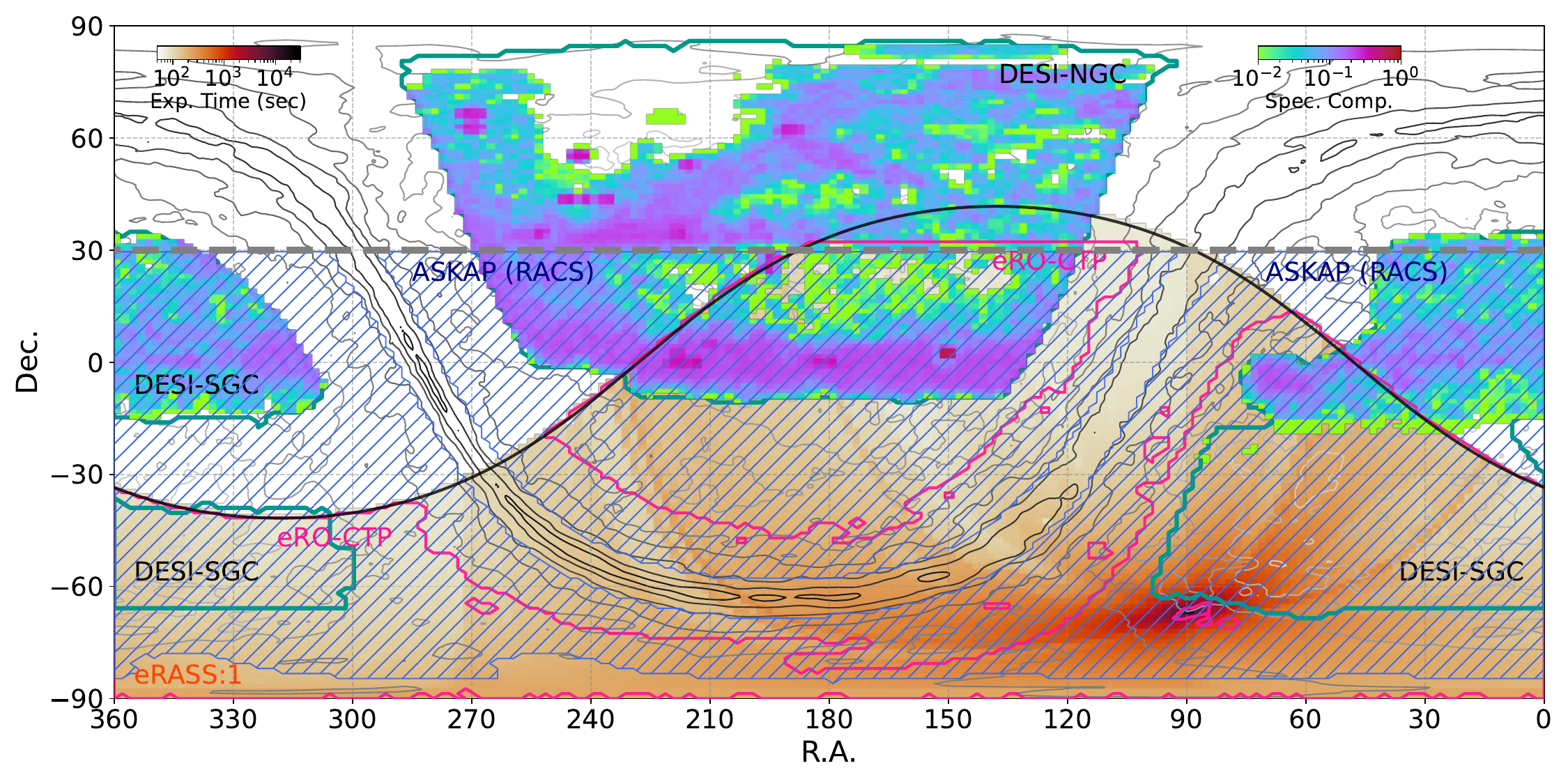}
    \caption{The sky coverages depicted in the DESI (enclosed by bold teal lines), ASKAP \citep[blue hatched region,][]{Hale..2021}, and eRASS1 (orange filled region), with the exposure times in the eRASS1 regions represented by the color bar. The regions enclosed by solid pink lines are the eRASS1 Counterpart Catalog \citep[][]{Salvato..2025, Kluge..2024}. The thin grey contour lines represent the galactic hydrogen column density along the line of sight to each point, based on HEALPIX resampling of the Leiden/Argentine/Bonn Survey of Galactic HI \citep{Kalberla..2005}. The filled neon contours display the spectroscopic completeness for galaxies in each region. The bold dashed grey line represents the region with $\delta < 30^{\circ}$, which corresponds to the area of data used in this study.}
    \label{fig:footprint}
\end{figure*}

\section{data}\label{sec:catalogs}

We start this section by detailing the galaxy and group samples from the DESI Legacy Surveys in Section~\ref{sec:desi}. Section~\ref{sec:spec} then discusses the characteristics of each galaxy. For an in-depth understanding of the DESI data release, refer to a collection of studies \citep[e.g.,][]{DESI..2016, DESI..2016b, DESI..2024, Levi..2013, Silber..2023, Miller..2024}. Furthermore, the Radio Catalog and X-ray catalog from eRASS1 will be presented in Section~\ref{sec:ctps}.

\subsection{DESI Legacy Imaging Survey Group Catalog}\label{sec:desi}

DESI Legacy Survey provides the currently largest photometry imaging in $g$/$r$/$z$ bands with 5$\sigma$ depth of 24.7/23.9/23.0 \citep{Dey..2019}. It consists of three independent optical surveys: the Beijing-Arizona Sky Survey (BASS), the Mayall $z$-band Legacy Survey (MzLS), and the DECam Legacy Survey (DECaLS). Although the raw data of the DESI Legacy Survey were obtained from different sets of photometric systems, they were reprocessed using the same pipeline and calibrated to be consistent with each other. The galaxy catalog also includes near-infrared data from the 6-year imaging of WISE, with a 5$\sigma$ depth of 20.7 and 20.0 in the 3.4 $\mu$m (W1) and 4.6 $\mu$m (W2) WISE bands. The galaxy catalog used in our study has removed the area within $|b| \leq 25^\circ$ to avoid regions of higher stellar density. The galaxy samples are selected with the morphology classification of \texttt{REX}, \texttt{EXP}, \texttt{DEV}, and \texttt{COMP} from Tractor fitting results. The photometric redshift of each galaxy is taken from the random forest algorithm-based  \textit{ photometric redshifts for the legacy surveys} \citep[PRLS,][]{ZhouRP..2021}, with a typical redshift error of $\sigma_z/(1+z) \sim 0.02$. Note that the reliability of photo-z estimation decreases beyond an apparent magnitude of the $z$ band of $m_z \simeq 21$ mag. Additionally, sources located within masked regions, such as those near bright stars, globular clusters, and large galaxies identified in the Siena Galaxy Atlas (SGA), have been excluded. The areas encompassing high Galactic-latitude supernova remnants (SNRs), such as SN1006 \citep{Winkler..2014}, Hoinga \citep{Becker..2021}, G296.5 + 10.0 \citep{Giacani..2000} and G330.0 + 15.0 \citep{Leahy..2020}, are not included within the DESI footprint.

The DESI group catalog used in this work is taken from \citet[][hereafter Y21]{Yang..2021}, which extended the halo-based group finder developed by \citet{Yang..2005, Yang..2007}. Briefly, the group finder starts by considering each galaxy within the redshift range of $0.0 < z_g < 1.0$ and $z-$band magnitude brighter than $m_{z} = 21$ mag as a group candidate. The cumulative group luminosity functions are then measured and the halo mass ($M_h$) is assigned to each candidate group using an abundance matching approach. From this, the halo radius ($R_{180}$) and line-of-sight velocity dispersion ($\sigma_{180}$) are estimated. Starting with the most massive group, members are re-identified in regions where the galaxy number density contrast exceeds a certain threshold. With the updated group memberships, both the center and total luminosity of the group are recalculated, and the process of measuring the luminosity function and assigning $M_h$ is repeated. The iteration continues until the mass-to-light ratios stabilize. This group finder was initially applied to the DESI Imaging Legacy Surveys DR8, while the present version has been updated based on the DR9 dataset.\footnote{DESI group/cluster catalogs are avaliable here: https://gax.sjtu.edu.cn/data/DESI.html}.

Note that each galaxy in the sample is assigned to a unique group, even if the group has only one galaxy member. Additionally, a small fraction of the redshifts have been replaced by spectroscopic data from the \texttt{FastSpecFit} Value-Added Catalogs \citep{Moustakas..2023} and other surveys \citep[see more details in][]{Yang..2021}. Consequently, the group catalog has been updated compared to the original version, containing $\sim 100$ million groups with $\sim 120$ million galaxy members with a sky coverage of $\sim 18500$ deg$^{2}$. Since eRASS1-DE only covers half of the sky, the sample size and the corresponding sky coverage used in this work are reduced ($\sim 43$ million groups). 

In Y21, the dark matter halos are defined as having an overdensity of 180 times the matter density of the Universe. The halo mass ($M_h$) of each group has been estimated based on the abundance matching between the total luminosity of the group and the halo mass assuming Planck18 cosmology \citep{Planck18}. The $M_h$ has an uncertainty of $\sim 0.2$ dex at the high mass end ($M_h \gtrsim 10^{14} h^{-1}M_\odot$), increasing to $\sim 0.4$ dex at $M_h \gtrsim 10^{12.3} h^{-1}M_\odot$, and then decreasing to $\sim 0.3$ dex at $M_h \gtrsim 10^{11} h^{-1}M_\odot$. The angular virial radius, $\theta_{180}$, is calculated using
\begin{equation}
    \theta_{180} = R_{180} \cdot D_A^{-1} = {\left(\frac{M_h}{\frac{4\pi}{3} \cdot 180\Omega_m \cdot \frac{3H_0^2}{8\pi G}}\right)}^{1/3} \cdot D_c^{-1},
\end{equation}
where $R_{180}$ is the virial radius, $D_A$ is the angular diameter distance, and $D_c$ is the comoving distance of that group.

\subsection{DESI Spectroscopic galaxy sample}\label{sec:spec}

We make use of the DESI spectroscopic observation data (up to the first year observation) from the \texttt{FastSpecFit} Value-Added Catalogs \citep{Moustakas..2023}. \texttt{FastSpecFit} analyzes the optical spectroscopy ($3600-9800 \mathring{A}$) and broadband photometry of extragalactic targets in the observed frame, utilizing physically based templates for stellar continuum and emission lines. The spectroscopic produced by the \texttt{Redrock} redshift-fitting pipeline \citep[][]{Guy..2023, Brodzeller..2023, Moustakas..2023, Schlafly..2023}, include best-fit redshift, emission line fluxes, stellar mass ($M_\star$), star formation rate (SFR), and the spectral classification based on the best-fitting template (SPECTYPE=STAR, GALAXY, QSO).

The DESI EDR and Year 1 spectra are reduced and released as `fuji', `iron', and `guadalupe' internal data releases, respectively. Combining the spectroscopic catalogs from these releases results in a considerable overlap of samples. Before prioritizing the best spectrum for objects with multiple spectra, we apply the \texttt{Redrock} fitting quality cuts: \texttt{ZWARN} = 0, to eliminate fiber issues and ensure reliable redshift measurements. If multiple spectra meet the aforementioned criterion, we then choose the one with the highest signal-to-noise ratio as the ``best spectrum" for the object. 

Consequently, there are approximately 2.5 million galaxy groups for which spectroscopic data have been obtained concerning their brightest group galaxies (BGGs). Including the satellite galaxies, this leads to a total of roughly 4.5 million galaxies. It is worth noting that while the BGGs in these galaxy groups have spectroscopic observations, this does not necessarily mean that all the corresponding satellite galaxies have complete spectroscopic measurements. Among $\sim 2$ million satellite galaxies, $\sim 190$k of them have spectroscopic measurements. However, the spectroscopic completeness of satellite galaxies does not affect our results. Due to the limited sky coverage of the X-ray counterpart and radio catalogs, groups in regions with $\delta > +30^{\circ}$ are excluded from consideration which will be discussed in detail in the next section. This results in a reduction of the number of groups with spectroscopic measurements for BGGs to $\sim 2.33$ million.

Since this work aims to understand the contribution of AGNs and warm/hot gas to the overall X-ray luminosity and profile, it is crucial to determine the $M_\star$ and star formation rate (SFR) of each galaxies. These parameters are essential to accurately calculate the contribution of X-ray binaries, and we refer to the model by \citet[][Model 5 in Table 2]{Aird..2017}. This empirical model provides luminosities in the $2-10$ keV band. We need to convert it to the X-ray luminosity in the $0.5-2$ keV band by assuming a power-law spectrum with a photon index of $1.8$. While \texttt{FastSpecFit} provide these quantities, a complete set of $M_\star$ and SFR is necessary to assess the impact of X-ray binaries from satellite galaxy on the overall X-ray profile of the X-ray binaries. Therefore, we also adopt the spectral energy distribution (SED) code \texttt{CIGALE} \citep[][]{Boquien..2019}, which is an alternative method for estimating $M_\star$ and SFR based on five-band photometry ($g$/$r$/$z$/W1/W2).

\subsection{X-Ray and Radio Catalogs}\label{sec:ctps}
For X-ray, we used the public eROSITA data from the eRASS1 field. The eRASS1 data are divided into 4,700 slightly overlapping sky tiles, each covering $3.6 \times 3.6$ deg$^2$. eROSITA-DE holds proprietary rights to regions in the Western Galactic hemisphere. The overlapping area between DESI and eRASS1-DE is $\sim 8700$ deg$^2$, $\sim 46 \%$ of the DESI footprint, as shown in the left panel of Figure~\ref{fig:footprint}. The eRASS1 data was processed with the eROSITA Science Analysis Software System (\texttt{eSASS}).

\subsubsection{eRASS1 Counterpart Catalogs}\label{sec:xmaps}
Following \citet{Merloni..2024}, we utilize \texttt{eSASS} on each tile to produce both the event list and its corresponding exposure map. The exposure times range from $\lesssim 200$ s to $\sim 40000$ s, spanning from the ecliptic equatorial region to the southern ecliptic pole. A considerable area near the southern ecliptic pole, which has the deepest exposure, is excluded because of the lack of optical coverage because of LMC. Compared with the depth curve of eFEDS, only a small region ($\sim 10$ deg$^2$) after cross-matching with DESI footprints has exposure times deeper than those of eFEDS. 

\citet{Merloni..2024} present two catalogs (main and supplementary) created in the $0.2 - 2.3$ keV band, containing $\sim 1.28$ million blind-detected X-ray sources with likelihood of detection $\mathcal{L}_{\rm det} \ge 5$. Most of these targets are point-like sources with likelihood of extent $\mathcal{L}_{\rm ext} = 0$, while only 26,682 of them with $\mathcal{L}_{\rm ext} > 3$ are more likely to be emitted from extended gas within massive groups \citep{Bulbul..2024}. This selection criterion differs slightly from the previous one applied to the eFEDS survey as described in \citet{Brunner..2022}, the updated criterion aims to retain high-redshift clusters and compact galaxy groups with angular sizes comparable to the PSF in the cluster catalog, thereby maximizing eROSITA's source discovery potential \citep[e.g.,][]{Bulbul..2022, Bulbul..2024}. 

Since some X-ray sources might be associated with our group samples, it is necessary to perform cross-matching before stacking. \citet{Salvato..2025} have identified the counterparts (CTP) to eRASS1 point sources using Gaia DR3 \citep[][]{Gaia..2023}, CatWISE2020 \citep[][]{Marocco..2021}, and DESI Image Legacy Survey DR10 \citep[][]{LiCH..2024} with
the Bayesian NWAY algorithm \citep[][]{Salvato..2019} and trained priors, while \citet{Kluge..2024} have matched the extended X-ray sources to the clusters of galaxies. In this work, we only adopt the point CTP matching results with the Legacy Survey DR10, while the extended CTP results follow those of \citet{Kluge..2024}. Briefly, when matching the X-ray sources from eRASS1 with optical sources, the NWAY algorithm first performs a coarse match within a 60'' search radius and then calculates the probabilities based on multiwavelength photometry. In the CTP catalog, most of the sources have photometric redshifts computed with the \texttt{Circlez} algorithm \citep{Saxena..2024}. These values might differ slightly from the photometric redshifts used in this work. Although about half of the redshifts deviate from our results by more than $|\Delta z| > 0.05$, such differences do not affect the outcomes provided by the NWAY algorithm. Therefore, we do not account for discrepancies arising from different photometric redshift estimates and directly adopt the matching results from the CTP catalogs. There are $\sim 100$k counterparts available in the full sample, while $\sim 11$k of them are available in the spectroscopic sample. Excluding sources located outside the footprint of the DESI, the majority of unmatched cases could be attributed to counterparts that are either high-redshift ($z > 1$) quasars or galaxy groups, or nearby ($z \lesssim 0.1$) extended galaxies that are masked by the DESI pipeline due to their large angular sizes. A minor fraction of the counterparts also fall below the $z-$band magnitude threshold.

\subsubsection{Radio Counterparts}\label{sec:radio}

To distinguish different types of AGN, we also cross-match the samples with the Rapid ASKAP Continuum Survey (RACS) catalog \citep[][]{Hale..2021}. This catalog was compiled from 799 tiles that could be convolved to a common resolution of $25''$, covering a large contiguous region in the declination range $\delta = -80^\circ$ to $+30^\circ$ with a median field rms sensitivity of 250 $\mu$Jy beam$^{-1}$ at 887.5 MHz. Our cross-matching was performed using a search radius of $5''$. Since the spatial resolution of one beam is $25''$, and the diameter of a beam corresponds to 5 pixels in the map, the positional error is therefore $5''$. If multiple targets are matched within this aperture, we assign the RACS source to the one with the largest $M_h$. However, such cases are extremely rare within our sample.

As shown in Figure~\ref{fig:footprint}, a subset of DESI groups within the eRASS1 footprint lies in the northern sky at declinations $\delta \ge +30^\circ$, a region that is not covered by either RACS or the eRASS1 Counterpart Catalog. To maintain sample completeness and consistency, we excluded all sources at $\delta \ge +30^\circ$ from both the full and spectroscopic samples. Consequently, the final sample comprises $\sim 40$ million groups, with spectroscopic redshifts available for the BGGs in $\sim 2.33$ million of them.

\begin{figure*}
    \centering
    \includegraphics[width=0.95\hsize]{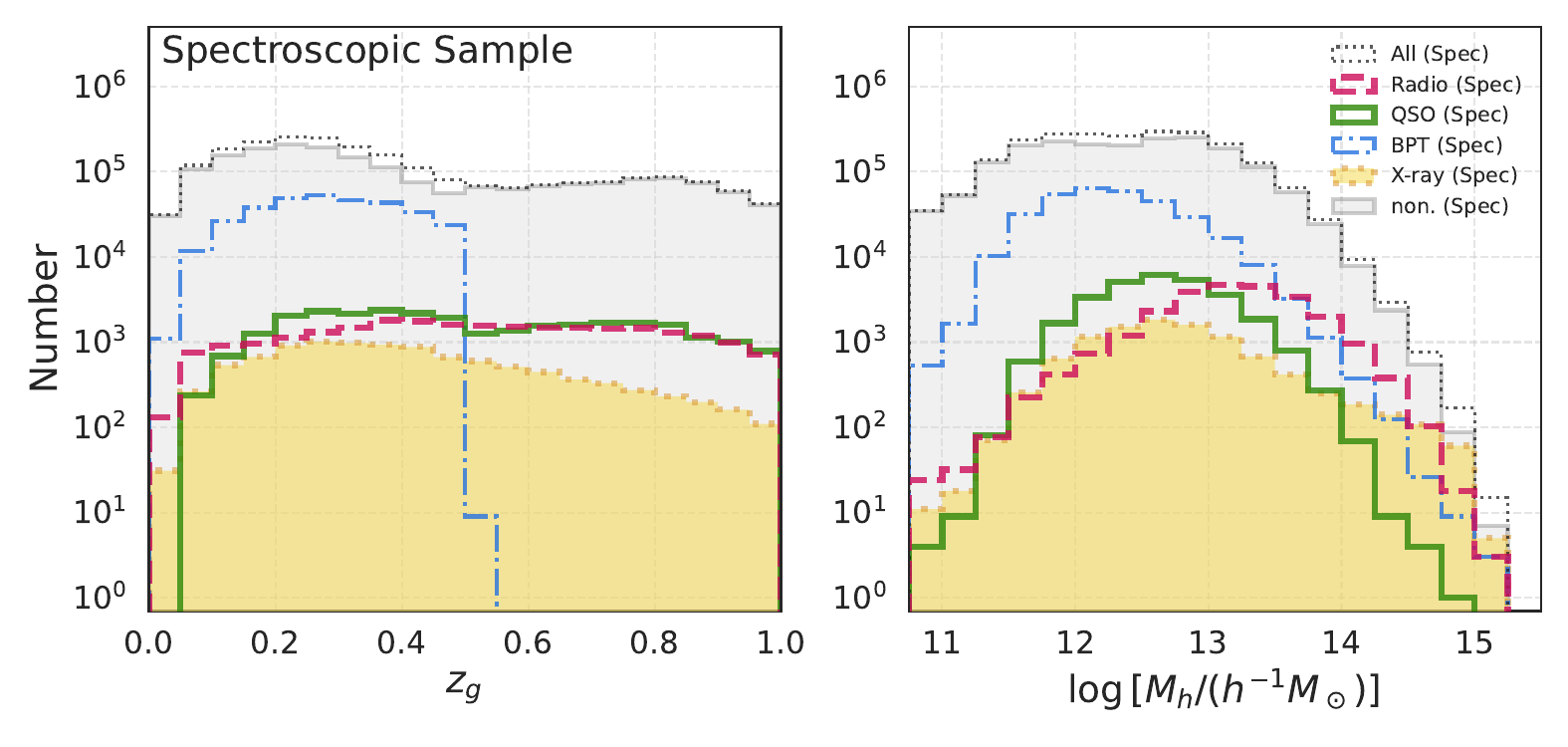}
    \includegraphics[width=0.95\hsize]{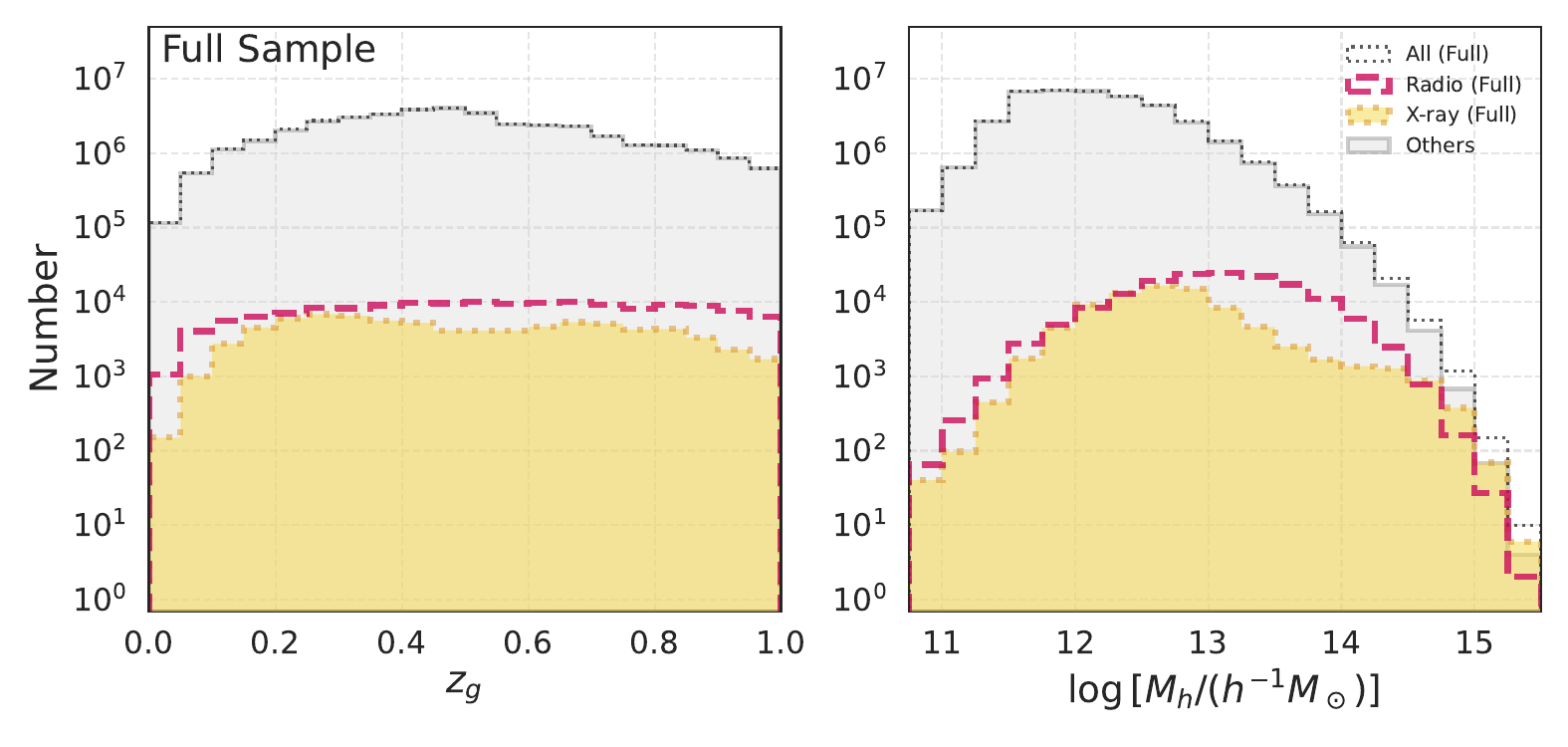}
    \caption{The distributions of $z_g$ (left) and $M_h$ (right) for the spectroscopic (upper) and full (lower) samples, respectively, categorized by their central characteristics: X-ray groups (yellow dotted and shaded), radio centrals (red dashed), QSO centrals (green solid), BPT-AGN centrals (blue dashdot), and non-AGN centrals or others (grey solid and shaded). Note that the full sample does not include systems with QSO or BPT-AGN centrals.}
    \label{fig:basicstat}
\end{figure*}

\begin{figure*}
    \centering
    \includegraphics[width=1\hsize]{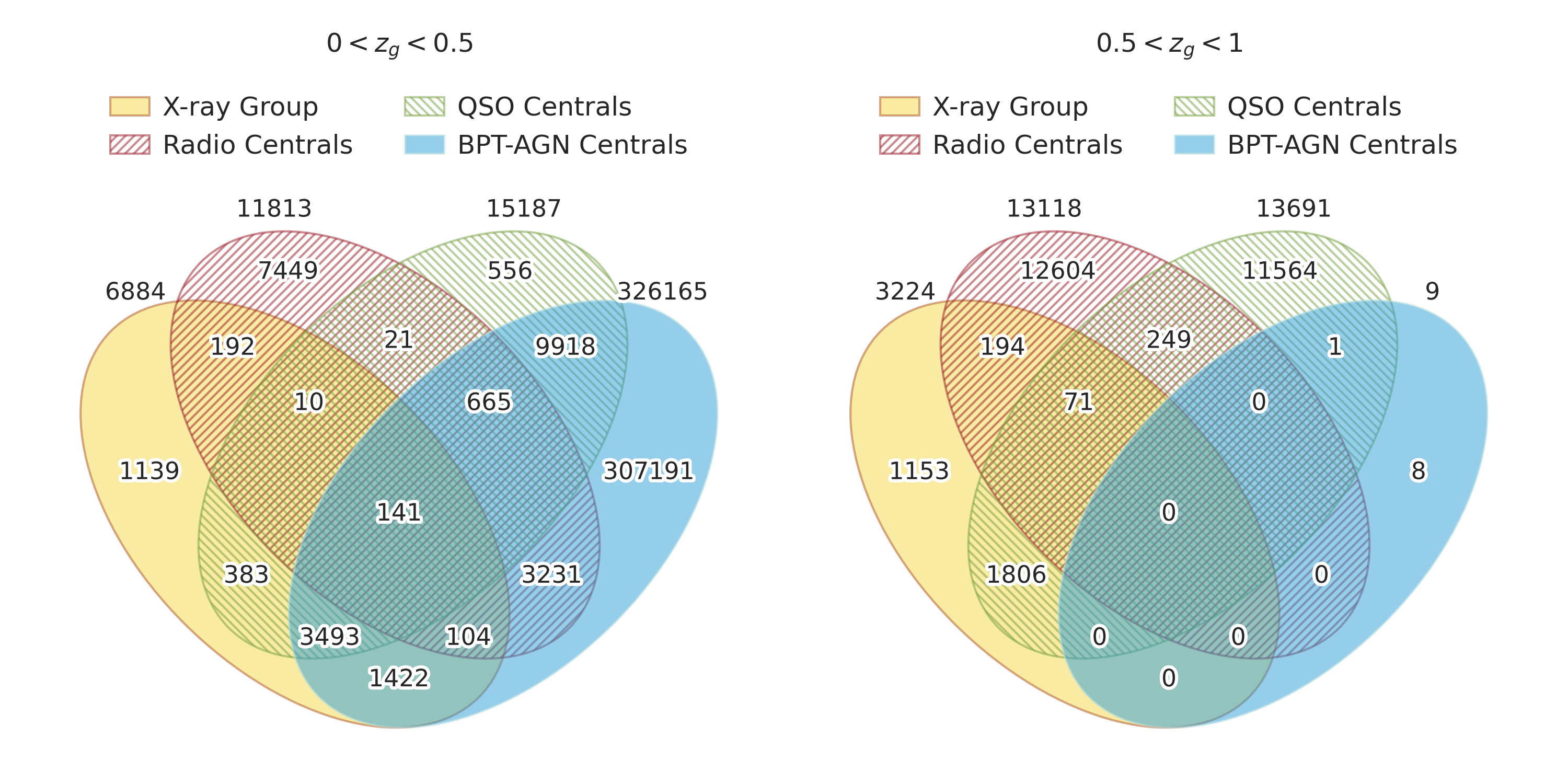}
    \caption{Venn diagrams illustrating the overlap among four AGN classifications: X-ray (yellow filled), Radio (red hatched), QSO (green hatched), and BPT-AGN (blue filled) for galaxy groups with BGGs having spectroscopic measurement. The left panel corresponds to groups with redshift $0 < z_g < 0.5$, while the right panel shows groups in the range $0.5 < z_g < 1.0$.}
    \label{fig:venn}
\end{figure*}

\begin{table*}[h!]
    \caption{Group types categorized by the AGN classification of their BGGs.}
    \centering\footnotesize
    \begin{tabular}{lcccl}
    \hline
    \hline
       \multirow{2}{*}{Classification} & Cross-matched Catalog & \multirow{2}{*}{Wavebands} & Maximum & \multicolumn{1}{c}{\multirow{2}{*}{Additional Notes}}\\
       & \footnotesize{(References)} &  & Separations & \\
    \hline
    \hline
       \multirow{2}{*}{X-ray Groups} & eRASS1 CTP & \multirow{2}{*}{X-ray} & \multirow{2}{*}{60''} & Select the unique one by combining the\\
       & \footnotesize{\citep[][]{Salvato..2025, Kluge..2024}} & & & astrometry and photometry.\\
    \hline
       \multirow{2}{*}{Radio Centrals} & Rapid ASKAP Continuum Survey & \multirow{2}{*}{Radio} & \multirow{2}{*}{5''} & Select the unique one by combining the \\
       & \footnotesize{\citep[][]{Hale..2021}} & & & $M_h$ of host halo.\\
    \hline
       \multirow{2}{*}{QSO Centrals} & \texttt{FastSpecFit} Value-Added & \multirow{2}{*}{Optical/NIR} & \multirow{2}{*}{0''} & This Catalog is a sub-sample of the\\
       & \footnotesize{\citep[][]{Moustakas..2023}} & & & Legacy Survey.\\
    \hline
       \multirow{2}{*}{BPT-AGN Centrals} & \texttt{FastSpecFit} Value-Added & \multirow{2}{*}{Optical/NIR} & \multirow{2}{*}{0''} & \multirow{2}{*}{As above}\\
       & \footnotesize{\citep[][]{Moustakas..2023}} & & & \\
    \hline
       \multirow{2}{*}{non-AGN Centrals} & \multirow{2}{*}{-} & \multirow{2}{*}{All} & \multirow{2}{*}{-} & Spectroscopic sample only, not X-ray, \\
       & & & & radio, QSO, or BPT-AGN. \\
    \hline
       \multirow{2}{*}{Others} & \multirow{2}{*}{-} & \multirow{2}{*}{All} & \multirow{2}{*}{-} & Full sample, not any of the above, but \\
       & & & & contain undetected QSOs and BPT-AGN. \\
    \hline
    \hline
    \end{tabular}
    \label{tab:classes}
\end{table*}

\section{Group types categorized by the AGN classification of their BGGs}\label{sec:class}
The aim of our study is to identify the sources that contribute to the X-ray emission in the galaxy groups. We need to distinguish different types of group based on the properties of the central galaxy. For simplicity, we have briefly listed the methods for groups categorized by the AGN classification of their BGGs in Table~\ref{tab:classes}.

Within the $\sim 40$ ($\sim 2.33$) million full (spectroscopic) samples, $\sim 82$k ($\sim 10$k) BGGs have X-ray counterparts, while $\sim 158$k ($\sim 25$k) have radio counterparts. Among them, $4,869$ ($565$) out of $\sim 82$k ($\sim 10$k) are extended X-ray sources. The X-ray emission associated with these groups might originate either from the BGG (point source) or from the hot gas (extended source). As \citet{Bulbul..2022} pointed out, a small fraction of the point sources are indeed under-luminous and compact in X-ray emission from lower-mass groups \citep[see also in][]{Balzer..2025}. Fundamentally, the difference between X-ray point sources and X-ray extended sources depends on multiple factors, including which component dominates the emission—AGN or hot gas—and the concentration of the hot gas itself, which also plays a significant role in the classification of X-ray sources. In fact, among the counterparts of these extended sources, $715$ out of $4,869$ are radio galaxies. For the spectroscopic sample, $17$ out of $565$ are QSOs and $42$ out of $565$ are BPT-AGN. This suggests that the extended X-ray component itself is somewhat associated with the activity of the central AGN. Therefore, in this work, any group with an X-ray counterpart is considered an X-ray group, regardless of whether it is a point source or an extended source. 

Furthermore, the identification of X-ray groups in this study differs from that in \citet{Zheng..2023}. In brief, in the previous paper we directly assigned each eFEDS-detected X-ray source to the most massive DESI group within an angular separation of $0.3R_{180}$, thereby leading to differences in the resulting fraction of X-ray groups (see Appendix~\ref{sec:im_stacking}). Although, as shown in Figure~\ref{fig:basicstat}, the fraction of X-ray groups is very low, especially for low-mass groups, this small subset is sufficient to affect the overall $L_{X}-M_{h}$ relation.

Radio centrals have already been introduced in Section~\ref{sec:radio}. As shown in Figure~\ref{fig:basicstat}, they are more likely to reside in higher-mass halos compared to other types of AGN centrals. This trend might be driven by two factors: first, radio galaxies might represent a later evolutionary stage of AGN activity; second, the relatively shallow depth of RACS might bias detections toward bright sources hosted by massive galaxies. \citet{Mandelbaum..2009} found that optical AGN reside in dark matter haloes with masses comparable to those of non-AGN galaxies at the same stellar mass, whereas radio-loud AGN tend to inhabit significantly more massive haloes at fixed stellar mass. \citet{Hickox..2009} found that X-ray-selected AGNs are preferentially hosted by galaxies located in the “green valley” of the color-magnitude diagram, typically residing in dark matter haloes of $M_h \sim 10^{13}h^{-1}M_\odot$, whereas radio AGNs are typically found in luminous red-sequence galaxies that inhabit more massive haloes of $M_h \sim 3 \times 10^{13}h^{-1}M_\odot$.

In contrast to the categories mentioned above, QSO and BPT-AGN centrals can only be applied to the subset of $\sim 2.33$ million galaxy groups with their BGGs having spectroscopic measurements. QSO Centrals are simply defined as galaxies whose SPECTYPE is classified as ‘QSO’. BPT-AGNs are identified using the BPT diagram \citep{Baldwin..1981}, which is traditionally used to distinguish type II AGNs from star-forming galaxies. We adopt the demarcation lines proposed by \citet{Kewley..2001} and \citet{Kauffmann..2003} for classification. In this work, the fluxes of four emission lines are required to have a signal-to-noise ratio of $S/N \ge 2$. Moreover, a significant portion of the samples that have their H$\alpha$ and $[\rm{N}\,\textsc{ii}]$ lines are detected with $S/N \ge 2$, but either or both H$\beta$ and $[\rm{O}\,\textsc{iii}]$ lines have lower $S/N$. Following \citet{CF..2010} and \citet{CF..2011}, we use the WHAN method, which only requires the equivalent width of $[\rm{N}\,\textsc{ii}]/\rm{H}\alpha$ and H$\alpha$ ($W_{\rm{H}\alpha}$) to select weak-line AGN. In order to maximize the identification of AGNs, we classify both composite and Seyfert galaxies in the two types of BPT diagrams as BPT-AGNs. Therefore, any group with spectroscopic measurements that does not fall into any of the aforementioned categories is classified as a non-AGN.

In Figure~\ref{fig:basicstat}, BPT-AGNs constitute a significantly larger fraction compared to the other categories mentioned above. The abrupt drop beyond $z_g \sim 0.5$ is due to the shift of the emission line $[\rm{N}\,\textsc{ii}]$ beyond $9800 \mathring{A}$, which is the maximum wavelength in the spectrum. 

There might be overlaps between different group classifications, some of which might be substantial. We examined the Venn diagram of these AGN classes within the $\sim 2.33$ million spectroscopic sample in Figure~\ref{fig:venn}. At lower redshifts ($z_g < 0.5$), a significant number of X-ray groups and QSO centrals are also identified as BPT-AGN centrals. Moreover, there is substantial overlap among X-rays, QSOs, and BPT-AGNs. In the regime where $z > 0.5$, there is considerable overlap between X-ray groups and QSO centrals. In contrast, radio galaxies exhibit little overlap with other types. The differing degrees of overlap among these subsamples are largely driven by the distinct environments in which various types of AGN reside. Although, for the BPT-AGN classification, the maximum identifiable redshift is limited to $z_g < 0.5$, giving a smaller volume in which to find higher-mass haloes, previous work by \citet{Mandelbaum..2009} has shown that radio AGNs and BPT-AGNs have mean halo masses of $1.6 \times 10^{13}h^{-1}M_\odot$ and $8 \times 10^{11}h^{-1}M_\odot$, respectively. In our sample, after excluding sources with $z_g > 0.5$, the median $M_h$ of radio centrals remains nearly the same as that over the full redshift range, around $M_h \sim 10^{13}h^{-1}M_\odot$, whereas the median $M_h$ for BPT-AGNs is about $M_h \sim 1.8 \times 10^{12}h^{-1}M_\odot$.

In this study, our focus is on the X-ray properties of galaxy groups hosting different types of central BGGs. Therefore, we allow overlaps between different categories. To maximize the sample size, X-ray groups and radio centrals are selected from the full sample, while QSOs, BPT-AGNs, and non-AGNs are restricted to the spectroscopic sample.

For groups with a confirmed X-ray counterpart, we adopt the X-ray central coordinates directly from the X-ray catalog, in order to minimize the potential overestimation of the extended component caused by offset effects. For those without an identified X-ray counterpart, the position of the BGG is used instead. As shown later, the BGG coordinates provide a good approximation to the center of the X-ray emission.

\begin{figure*}
    \centering
    \includegraphics[width=1.\hsize]{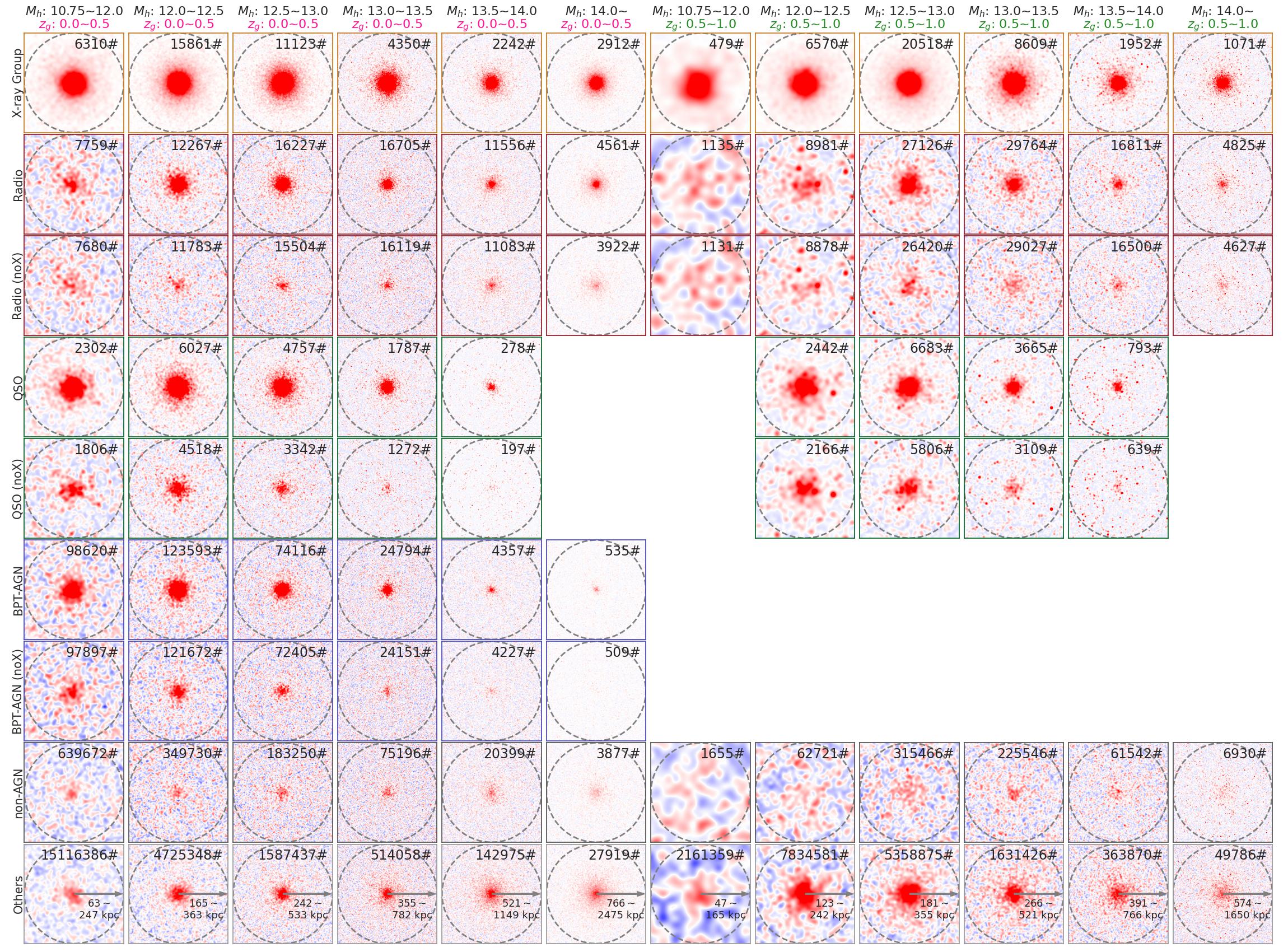}
    \caption{Stacked rest-frame $0.5-2.0$ keV band images of DESI groups with different central properties at different $M_h$ and $z_g$ bins. The dashed circles represent the regions within a radius of $R_{180}$. Only the data bins with at least 100 groups are shown here.The bottom row of each column indicates the radial scale range corresponding to the respective $M_h$ and $z_g$ bins. The number of subsamples for each category within a given bin is shown in the top-right corner of each panel. }
    \label{fig:xstack}
\end{figure*}

\begin{figure*}
    \centering
    \includegraphics[width=.495\hsize]{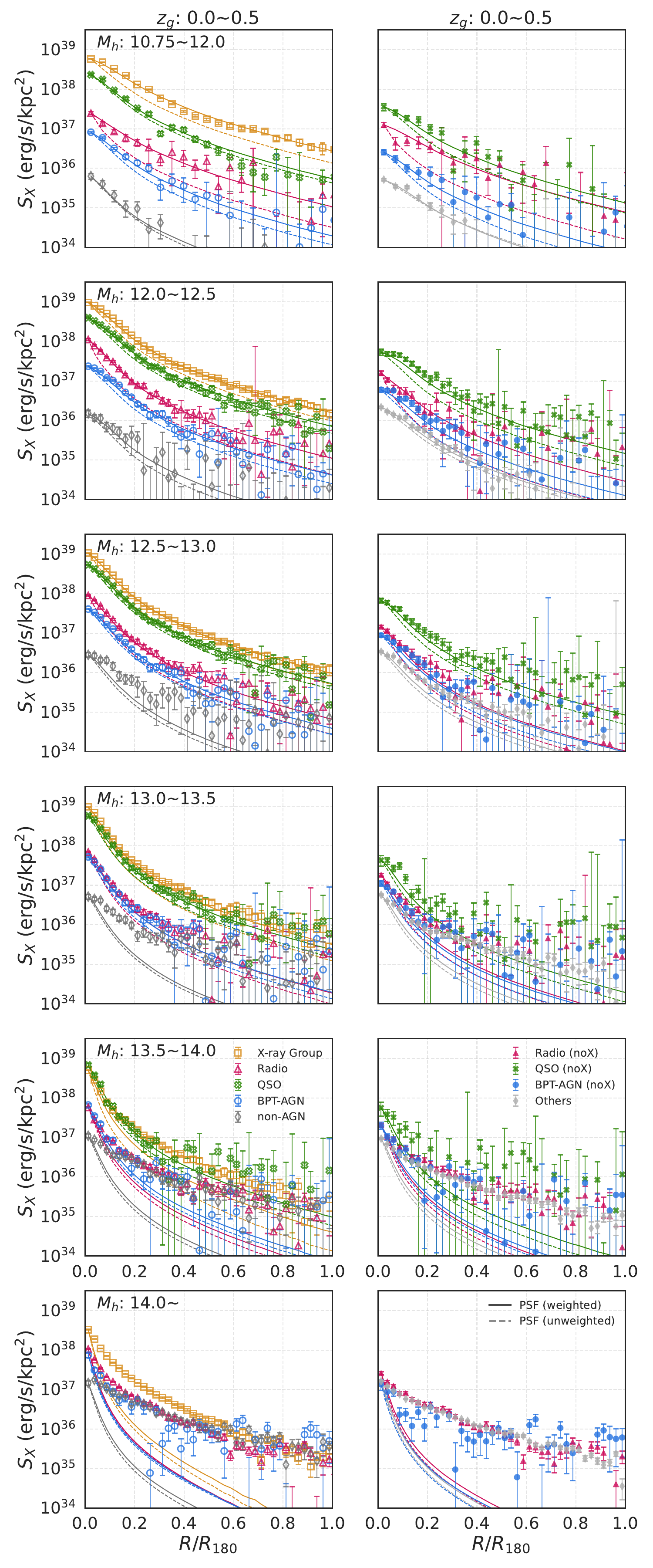}
    \includegraphics[width=.495\hsize]{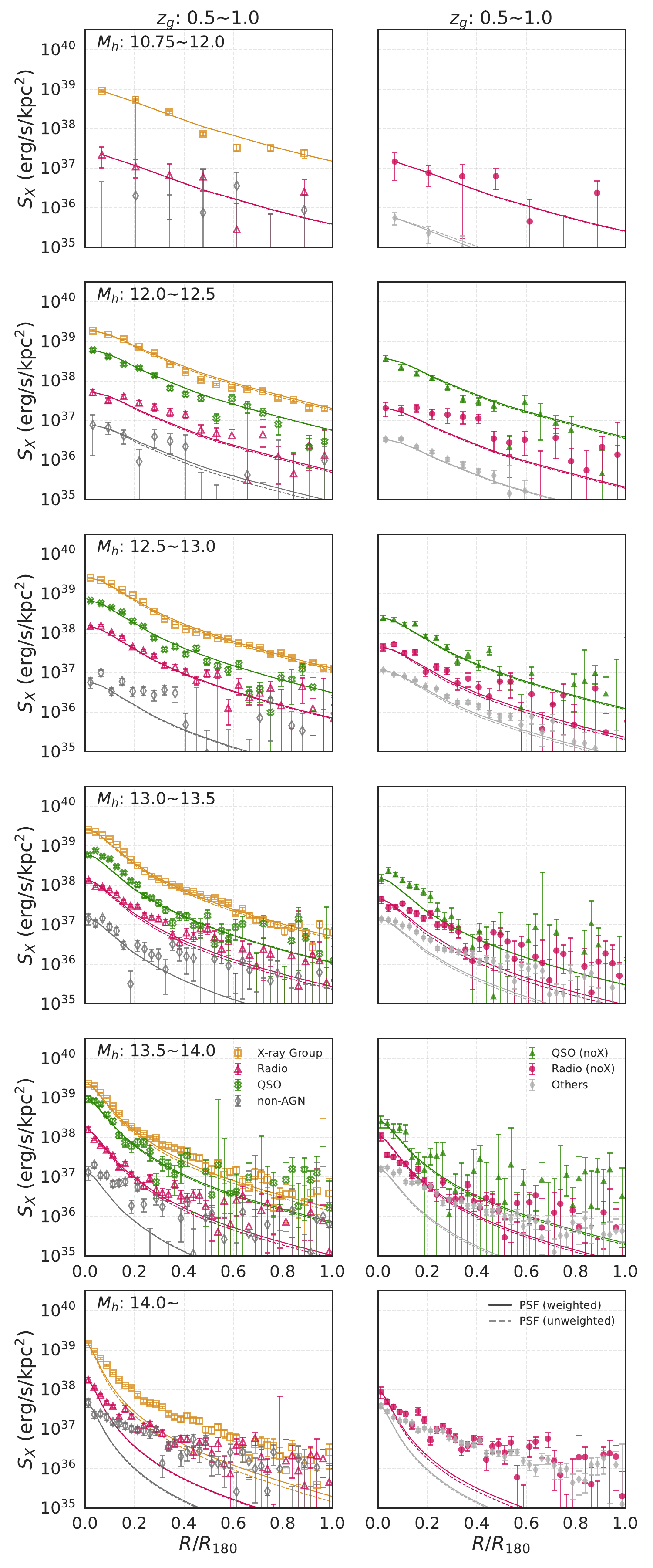}
    \caption{Left two columns: Mean X-ray surface brightness profiles of $z_g < 0.5$ DESI groups with different central properties at different $M_h$ bins. The leftmost corresponds to different catagories, whereas the rightmost of the two left columns shows the subsamples after removing the X-ray detected ones. The solid lines represent the convolved PSF normalized by their individual count rates, while the dashed lines represent the convolved PSF normalized without weighting. Only the data bins with at least 100 groups are shown here. Right two columns: The same as the left two columns, but for the groups with $z_g > 0.5$.}
    \label{fig:xprofile}
\end{figure*}

\begin{figure}
    \centering
    \includegraphics[width=1.\hsize]{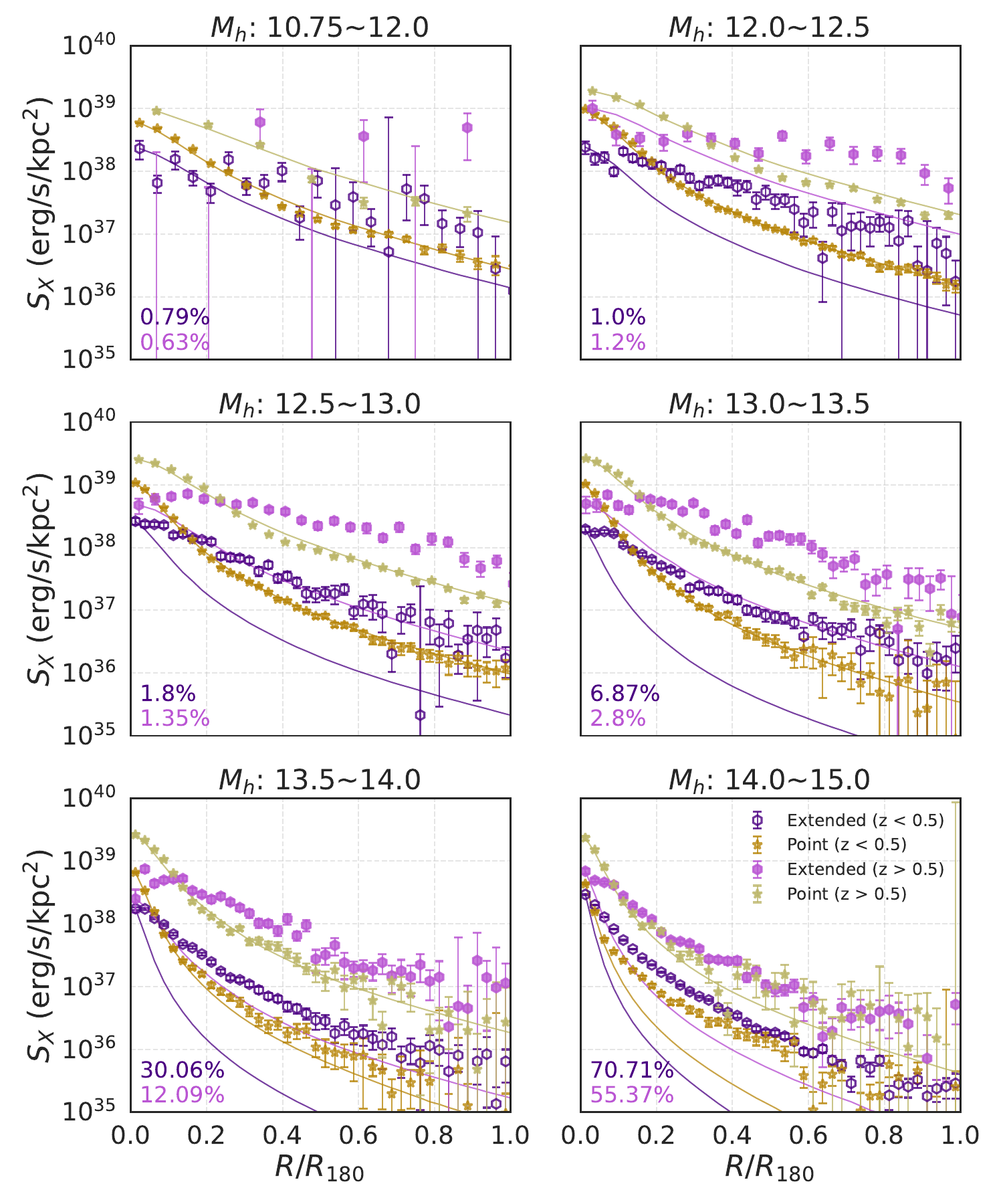}
    \caption{Mean X-ray surface brightness profiles of X-ray groups identified as extended (purple) and point (gold) sources in the eRASS1 pipeline for different $z$ and $M_h$ bins. The solid lines represent the convolved PSF normalized by their individual count rates. In each panel, we highlight the fraction of extended groups relative to all X-ray groups in each $M_h$ bin.}
    \label{fig:xxprof}
\end{figure}

\section{Results} \label{sec:results}
Since the X-ray measurements and stacking procedures are rather complicated, involving extensive details of X-ray profiles and PSF convolution, we place all details to the Appendices. In Figure~\ref{fig:xstack}, we present the stacked images of the groups categorized by their central characteristics. Only data bins containing at least 100 groups are shown. It is clear that all group types exhibit a central X-ray excess. Of course, many of these types include X-ray-detected groups, but even after removing those with X-ray detections from the Radio, QSOs, and BPT-AGN categories, a clear X-ray excess remains visible. From the stacked images, it is also evident that the X-ray brightness varies among different types of group. In this section, we will delve deeper into comparing their X-ray profiles and luminosities.

\subsection{X-Ray Surface Brightness Profile}

In Figure~\ref{fig:xprofile}, we illustrate the combined X-ray surface brightness profiles for groups exhibiting varying central characteristics across different $M_h$ and $z_g$ intervals. Notably, the X-ray signals are markedly enhanced post-stacking, even for the subsets where the X-ray detected ones have been excluded.

When considering the overall X-ray surface brightness, X-ray groups are inherently the most prominent, with QSOs, radio sources, BPTs, and non-AGNs following in that order. To be more specific, the X-ray brightness of X-ray groups and QSOs are quite similar. As shown earlier in Figure~\ref{fig:venn}, the overlap between these two categories is very high. Meanwhile, the difference between radio and BPT-AGNs is also very small, especially in more massive halos. The general trend is that as $M_h$ decreases, the differences between the sub-samples become more pronounced. However, in the most massive bins, the differences in the profiles of the outskirts are less noticeable, except for the central excess.

In fact, the X-ray profile might potentially be caused by the point spread function (PSF) of the central point source\footnote{The PSF in the CALDB of eROSITA is still from the gound calibration efforts and might not reflect the final PSF of eROSITA.}, so we also plot the convolved PSF models in Figure~\ref{fig:xprofile} (see details in Appendix~\ref{sec:psf}). In summary, during the PSF convolution, we considered two models: the first assumes a consistent luminosity for the central point source across each subsample, and the second assigns weights based on the specific luminosity within $30''$ of each group.

It is clearly evident that the latter model fits the X-ray profile better than the former, especially for X-ray groups with $M_h \lesssim 10^{13.5}h^{-1}M_\odot$. This indicates two key observations: first, the model that weights luminosity offers a better depiction of the central point source's contribution; second, low-mass groups are significantly influenced by the PSF, especially in X-ray and QSO central groups, due to the exceptionally high brightness of their AGN emission. Since extended X-ray sources are rare in lower $M_h$ bins, low-mass X-ray groups are mostly point sources given by the eRASS1 pipeline. However, for X-ray groups with $M_h \gtrsim 10^{13.5}h^{-1}M_\odot$, the outer region of their surface brightness profile becomes significantly higher than the convolved luminosity-weighted PSF. 

The most direct reason is that in the massive $M_h$ bins, the fraction of extended X-ray sources increases. As noted in each panel of Figure~\ref{fig:xxprof}, the fraction of extended groups in X-ray groups for $z_g < 0.5$ shows a significant increase starting from $M_h \gtrsim 10^{13.5}h^{-1}M_\odot$,  while for $z_g > 0.5$, the increase becomes noticeable only at $M_h \gtrsim 10^{14}h^{-1}M_\odot$ due to the dimming of X-ray signal. Nevertheless, even when isolating the X-ray point sources within the most massive $M_h$ bin, the extended component in the outer parts of their X-ray profiles is still evident. As shown in Figure~\ref{fig:xxprof}, for nearby X-ray groups with $z_g < 0.5$, a slight excess in the outer regions of point X-ray sources begins at $M_h \gtrsim 10^{13.5}h^{-1}M_\odot$, and becomes quite evident at $M_h \gtrsim 10^{14}h^{-1}M_\odot$. For distant X-ray groups with $z_g > 0.5$, a similar excess in the outer regions of the profile is also noticeable at $M_h \gtrsim 10^{14}h^{-1}M_\odot$. This suggests that the difference between X-ray point sources and X-ray extended sources might stem from the relative contribution of AGN as point sources versus extended gas, or it could be caused by a higher concentration of the extended hot gas.

Indeed, for sources presenting a fairly strong point source, the overall profile bears a closer resemblance to the PSF when weighted by luminosity. Groups with QSO centrals, which have surface brightness comparable to that of X-ray groups, exhibit a similar characteristic. In contrast, groups with relatively weaker central components, such as radio, BPT-AGN centrals, and non-AGN groups, show excess emission in the outer regions even at $M_h \sim 10^{12.5}h^{-1}M_\odot$. On the other hand, incorporating X-ray extended sources within these subsamples could readily amplify the outer areas of the X-ray profile for those subsamples characterized by generally weaker X-ray emission. We attempted to remove the X-ray detected subset from the QSO, Radio, and BPT-AGN central groups to prevent X-ray blind-detected groups from having an overly decisive effect. However, the excess in the outer regions becomes more pronounced. Moreover, the relative brightness ranking among them remains unchanged.

In summary, a general trend is observed in which the extended X-ray emission in the outskirts becomes more prominent in systems with either more massive $M_h$ or lower total X-ray luminosity. However, whether the luminosity of the extended gaseous component varies across different classifications remains unclear from the X-ray surface brightness profiles alone. This constitutes a key question to be addressed in the next section, and this would be the focus of discussion in the following section.

\begin{figure*}
    \centering
    \includegraphics[width=.95\hsize]{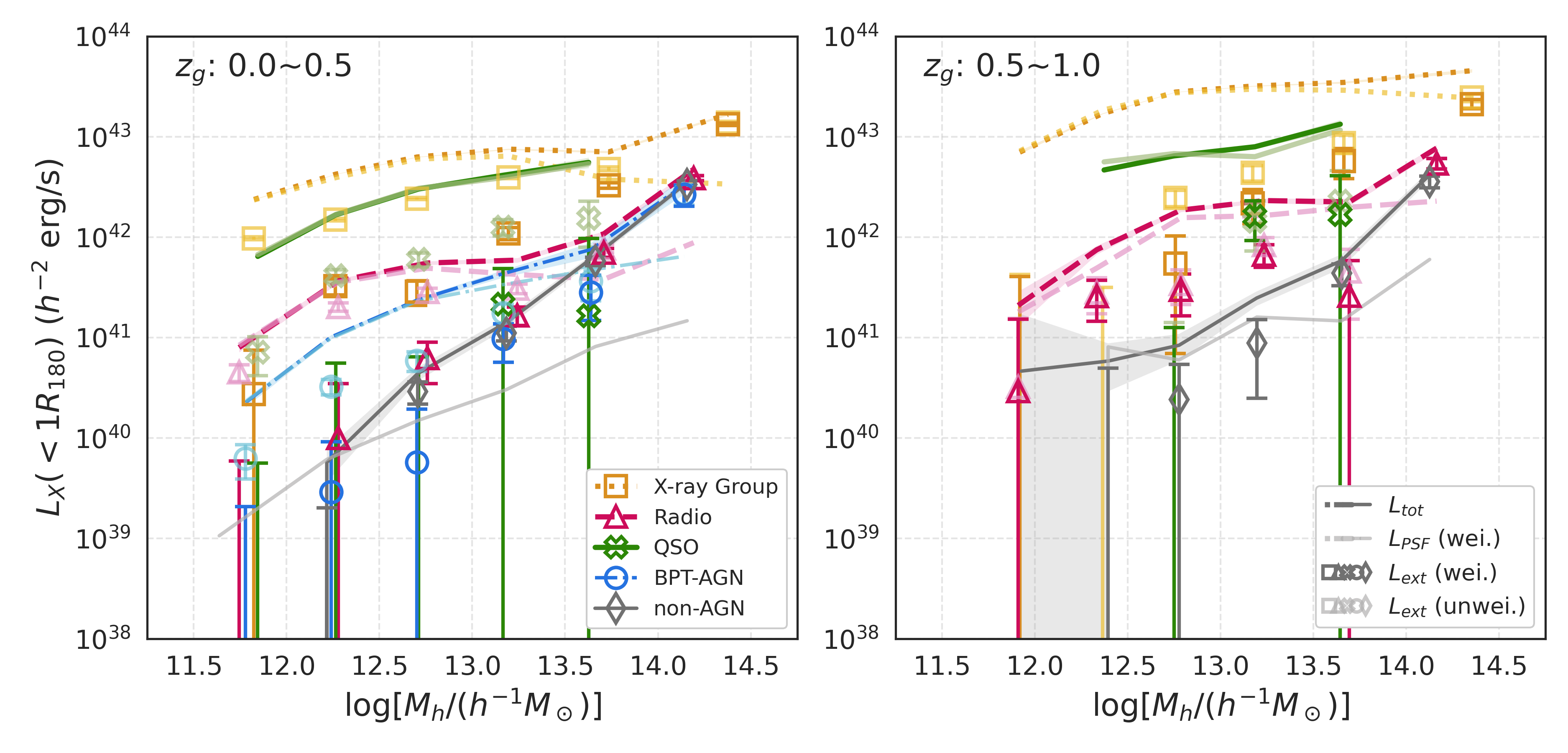}
    \includegraphics[width=.95\hsize]{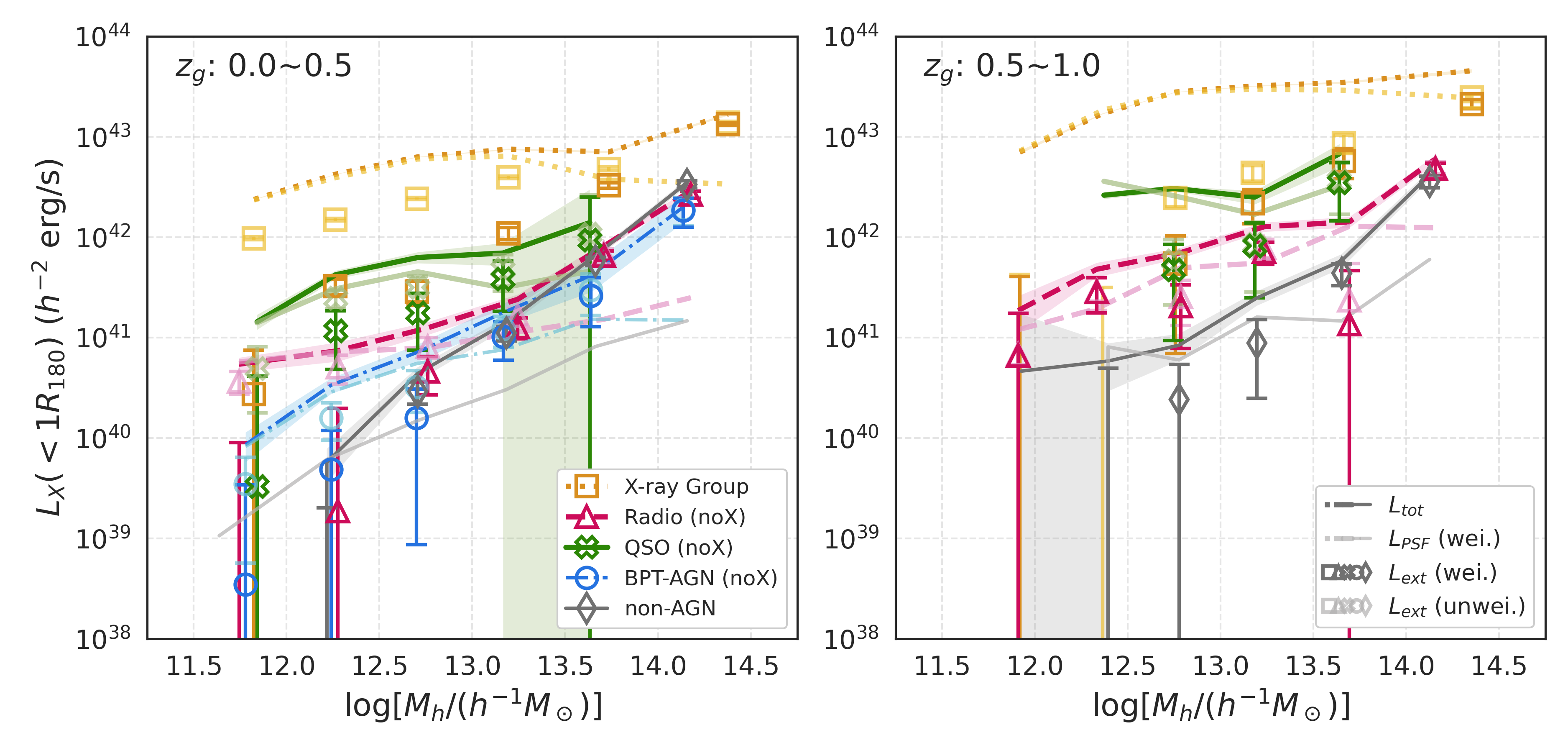}
    \caption{The rest-frame $0.5-2.0$ keV luminosity, $L_X$, versus halo mass, $M_h$, for groups with different central properties, including X-ray groups (yellow dotted), QSOs (green solid), Radio (red dashed), BPT-AGN (blue dashdot), and non-AGN groups (grey solid). Note that the groups with QSO, BPT-AGN, and non-AGN centrals are drawn from the spectroscopic sample, while the others are drawn from the full DESI samples. In the upper panels, each subsample includes all objects that meet the selection criteria, whereas the lower panels show the same subsamples but with Radio, QSOs, and BPT-AGNs after excluding those that belong to X-ray groups. Dark lines with shaded regions represent the total $L_X$, while semi-transparent lines indicate the convolved luminosity-weighted PSF component. Dark symbols with error bars denote the extended-gas luminosity after subtracting the luminosity-weighted PSF, while semi-transparent symbols with error bars show the result after subtracting the unweighted PSF.}
    \label{fig:LX3}
\end{figure*}

\begin{figure}
    \centering
    \includegraphics[width=1.\hsize]{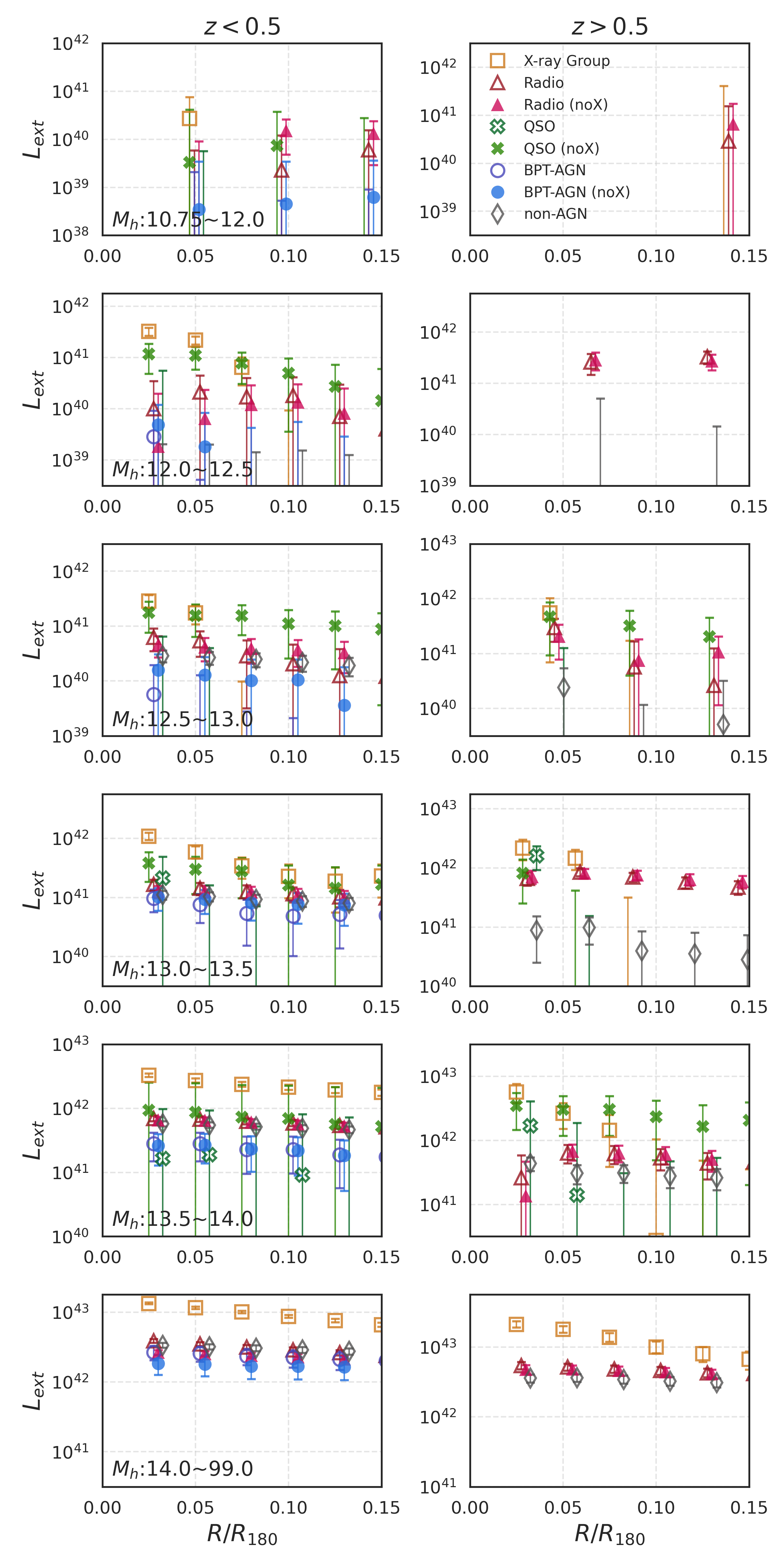}
    \caption{The X-ray emission from the gas component after subtracting the central PSF component normalized using different central radial ranges for different types of group. The data points have been manually offset along the x-axis for clarity.}
    \label{fig:Normalize}
\end{figure}

\subsection{X-ray Luminosity}
The X-ray luminosity of galaxy groups primarily consists of three components: AGN, hot gas, and X-ray binaries. By decomposing these contributions (see details in Appendix B), we present in Figure~\ref{fig:LX3} the rest-frame $0.5-2.0$ keV X-ray luminosity for the total emission, the PSF component, and the extended gas component as a function of $M_h$ across different group categories. 

Across all group categories, we observe a consistent positive correlation between $M_h$ and $L_X$. When directly comparing the total X-ray luminosities of different group categories at the same redshift and $M_h$, we find the following hierarchy in total $L_X$: X-ray Groups, QSO, Radio, BPT-AGN, and non-AGN centrals. However, the mass dependence shows a clear category dependence, where group types with relatively weaker $L_X$ show steeper slopes. This results in amplified luminosity contrasts, particularly in lower $M_h$ regimes, where the differences become more pronounced.

After separating the central PSF component, it becomes clear that the central PSF part of low-mass groups is almost identical to the total $L_X$. For groups with $z_g < 0.5$, we can distinctly isolate the central PSF component from the total $L_X$ only when $M_h \gtrsim 10^{13.5}h^{-1}M_\odot$. For $z_g > 0.5$, only the groups within the most massive categories can somewhat distinguish the PSF component effectively. There are several reasons behind this result. First, if the angular size, $\theta_{180}$, is too small to be comparable with the PSF size and the count rates are insufficient, it becomes very difficult to separate the PSF component from the overall brightness profile of the X-ray surface. For example, groups with $z_g > 0.5$, unless they are particularly massive, would suffer from this problem. Secondly, if the total $L_X$ of a group is overly dominated by the central AGN, it becomes challenging to disentangle the extended gas component. As illustrated in the upper left panel of Figure~\ref{fig:LX3}, non-AGN groups, which exhibit the lowest X-ray luminosity, can effectively isolate the extended gas component for the same $M_h$, specifically when $M_h \gtrsim 10^{12.5}h^{-1}M_\odot$. It is also clear from the X-ray surface brightness profile that radio centrals, where the central AGN is not particularly dominant, can stably separate the extended gas component, even in the $z_g > 0.5$ regime. Conversely, groups centered around QSOs often struggle to detach the extended component because of the predominance of X-ray point sources and the scant presence of X-ray extended sources, resulting in an overfocus on the central AGN. In contrast, for X-ray groups with $z_g < 0.5$, the relative error regarding the extended gas component is less pronounced. Although the proportion of X-ray extended groups is not high for $M_h \lesssim 10^{13}h^{-1}M_\odot$, it remains adequate to impact excess in the outer regions.

It is worth noting that when calculating the PSF component, we normalize the brightness at the center of the convolved PSF profile to the overall X-ray surface brightness profile. The choice of central radius might significantly affect the results. In this work, a smaller radius is preferable for this normalization, but insufficient pixel sampling would lead to significant uncertainties. Conversely, if the selected region is too large, it might lead to unrealistic results. The ideal scenario is when the outer regions of the surface brightness are much higher than the convolved PSF, as this would lead to the most robust results. In Figure~\ref{fig:Normalize}, we plot the X-ray emission of the extended gas component after subtracting the normalized central PSF component using different central radial ranges for different types of groups. It is evident that the groups with QSO centrals exhibit extremely large error bars, making the results highly unstable and not robust. In contrast, the results for groups with radio centrals are much more stable and reliable.

Therefore, to obtain more robust results, we excluded the subset containing X-ray detected sources from the groups with radio, QSO, and BPT-AGN centrals. In the bottom panels of Figure~\ref{fig:LX3}, although the average total $L_X$ decreases after removing the X-ray detected overlapping subsets, their relative order remains consistent. However, in the bins where $z_g < 0.5$ and $M_h \gtrsim 10^{13}h^{-1}M_\odot$, the differences between non-X-ray groups with radio, BPT-AGN, and non-AGN central are minimal, with non-AGN central groups even surpassing non-X-ray groups with BPT-AGNs in the most massive bin. Additionally, separating the luminosity of their extended gas component is easier, as indicated by their reduced relative errors. Especially for non-X-ray groups with QSO central, as shown in Figure~\ref{fig:Normalize}, we easily see that the robustness of the extended gas component luminosity for non-X-ray groups with QSO central improves significantly. In contrast, the impact on radio central groups after removing the subset containing X-ray detected sources is not that pronounced. 

When considering the X-ray luminosity for extended gas, it is clear that the X-ray groups are the brightest, followed by the groups without X-ray detection in eRASS1 with QSO centrals, though the difference between the two is not very significant. In contrast, the non-X-ray detected groups with radio, BPT-AGN, and non-AGN centrals are noticeably weaker, though the relative differences among these three categories are not very large. Although selection effects are involved here, and the X-ray selection is redshift and surface-brightness dependent, making it difficult to draw conclusions from comparisons of differently selected samples, in this case removing the X-ray-detected overlaps is intended solely to better highlight the extended X-ray component. Combined with the fact that groups with QSO centrals contain a higher fraction of X-ray groups compared to Radio and BPT-AGN centrals, we have reason to suspect that this result reflects an underlying coevolution between AGN activity and the extended gas component within the halo. Following the model proposed by \citet{Churazov..2005}, the bolometric luminosity of central supermassive black holes (SMBHs) strongly correlates with their accretion rates. Our results suggest that the central SMBH accretion rate might intrinsically linked to the X-ray luminosity of the surrounding gaseous component. This trend becomes increasingly pronounced in lower-mass groups, as their smaller virial radius makes the entire halo more susceptible to AGN feedback.

So far, we can conclude that for low-mass groups, the difference in $L_X$ among different categories of groups is primarily due to the discrepancy in the PSF component. It is only when $M_h \gtrsim 10^{13.5}h^{-1}M_\odot$ that extended gas starts to play a significant role. If we examine the relationship between PSF luminosity and $M_h$, it becomes evident that as $M_h$ increases, the relative difference in PSF luminosity decreases slightly, and there still remains a noticeable order of magnitude difference. This indirectly suggests that only in massive groups does the frequent collision of dark matter halos lead to gas heating, making it X-ray visible.

It is crucial to acknowledge that the findings we presently have are founded on an implicit assumption: regardless of whether all the X-ray emission from a group comes from its central galaxy, it is presumed that the optical location of the central galaxy exactly coincides with its position on the X-ray map. In contrast to this assumption, if positional offset between the peak of the X-ray emission and the BGG are frequently observed across a large number of groups, it becomes necessary to consider how such an effect could impact the convolved PSF and subsequently alter the estimation of the extended gas component. This topic will be explored in depth in the upcoming section.
\begin{figure}
    \centering
    \includegraphics[width=1.\hsize]{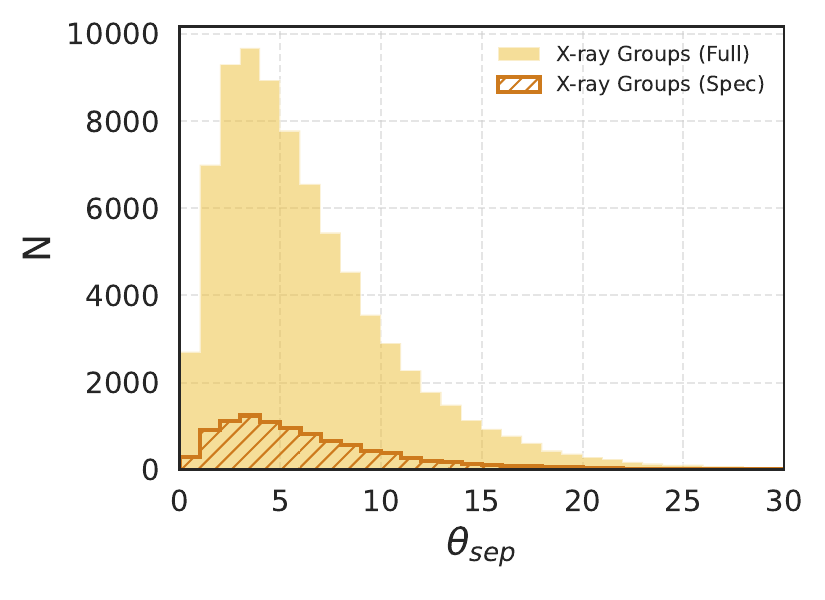}
    \caption{The distribution for angular separation between optical and X-ray positions for X-ray groups.}
    \label{fig:xsep}
\end{figure}

\section{Discussion}\label{sec:discuss}
In section~\ref{sec:results}, we demonstrated that $L_X$ varies across different categories of groups, with the variations being more significant in the low-mass range. Additionally, the non-X-ray groups featuring QSO centrals, known for their highest total $L_X$, possess an extended gas component that is evidently brighter compared to those in other groups, like radio, BPT-AGN, and non-AGN centrals, whose total $L_X$ is relatively less intense. 

As we mentioned in the previous section, positional offset between the peak of the X-ray emission and the BGG could lead to uncertainties in the findings. The observed offset might be attributed to either an intrinsic separation or artifacts introduced by measurement uncertainties. 

As suggested by \citet{Balzer..2025}, significant offsets between X-ray emission and BGG are typically associated with dynamically disturbed systems. Moreover, efficient accretion onto the SMBH of the central galaxy is known to impact the gas on very large scales within the dark matter halo \citep[][]{Eckert..2021}. The central AGN not only drives large-scale reorganization of the gas, but also generates significant morphological diversity through jets that excavate cavities in the intracluster medium. This feedback can substantially disturb the gas structure by expelling material from the center of the group \citep[e.g.,][]{Gaspari..2012, Gitti..2012, McNamara..2012, Li..2015, CuiWG..2016, Bahar..2024}.

On the other hand, systematic positional discrepancies between the X-ray and optical counterparts are unavoidable, even with more in-depth exposures and improved flux sensitivity \citep[e.g.,][]{Seppi..2023, Merloni..2024}. Should these undetected X-ray sources be observed by eROSITA through deeper exposure, their locations typically do not precisely match those given by optical surveys. This shift effect could increase the PSF width, leading to a decrease in luminosity contribution from the extended gas component, which might be initially overestimated \citep[e.g.,][]{Clerc..2024}. 

It remains challenging to quantitatively distinguish how much of the observed offset is attributable to intrinsic gas displacement induced by AGN feedback, and how much results solely from observational uncertainties. In this work, we consider only the most extreme scenario in which all observed offsets are assumed to result from systematic observational errors. Under this assumption, an artificial offset should be introduced when convolving the PSF. If a significant fraction of the groups actually exhibit a genuine displacement between the X-ray peak and the BGG due to AGN feedback, then the separation between the PSF center and the BGG would be intrinsically smaller. In other words, in this extreme case, where all PSFs are convolved with the assumed offset, the resulting broadening of the PSF profile represents the upper limit of this effect, and the residual gas component after PSF subtraction corresponds to the lowest possible estimate. Figure~\ref{fig:xsep} illustrates that the distribution of angular separation between optical and X-ray positions for the X-ray groups peaks near $3''$, with most sources found within $10''$. Despite being minor, this offset can significantly influence the outskirts.

A straightforward approach to assess this impact is to randomly displace the PSF center from the BGG's optical position when convolving the PSF. This random displacement follows the distribution of angular separations between the optical and X-ray positions for the X-ray groups. Note that this test is only applicable to subsamples without X-ray blind detection since the positions of the X-ray groups are directly provided by the CTP catalog. Therefore, we can only test the extended gas component of non-X-ray groups with QSO, radio, BPT-AGN, and non-AGN centrals.

\begin{figure}
    \centering
    \includegraphics[width=1\hsize]{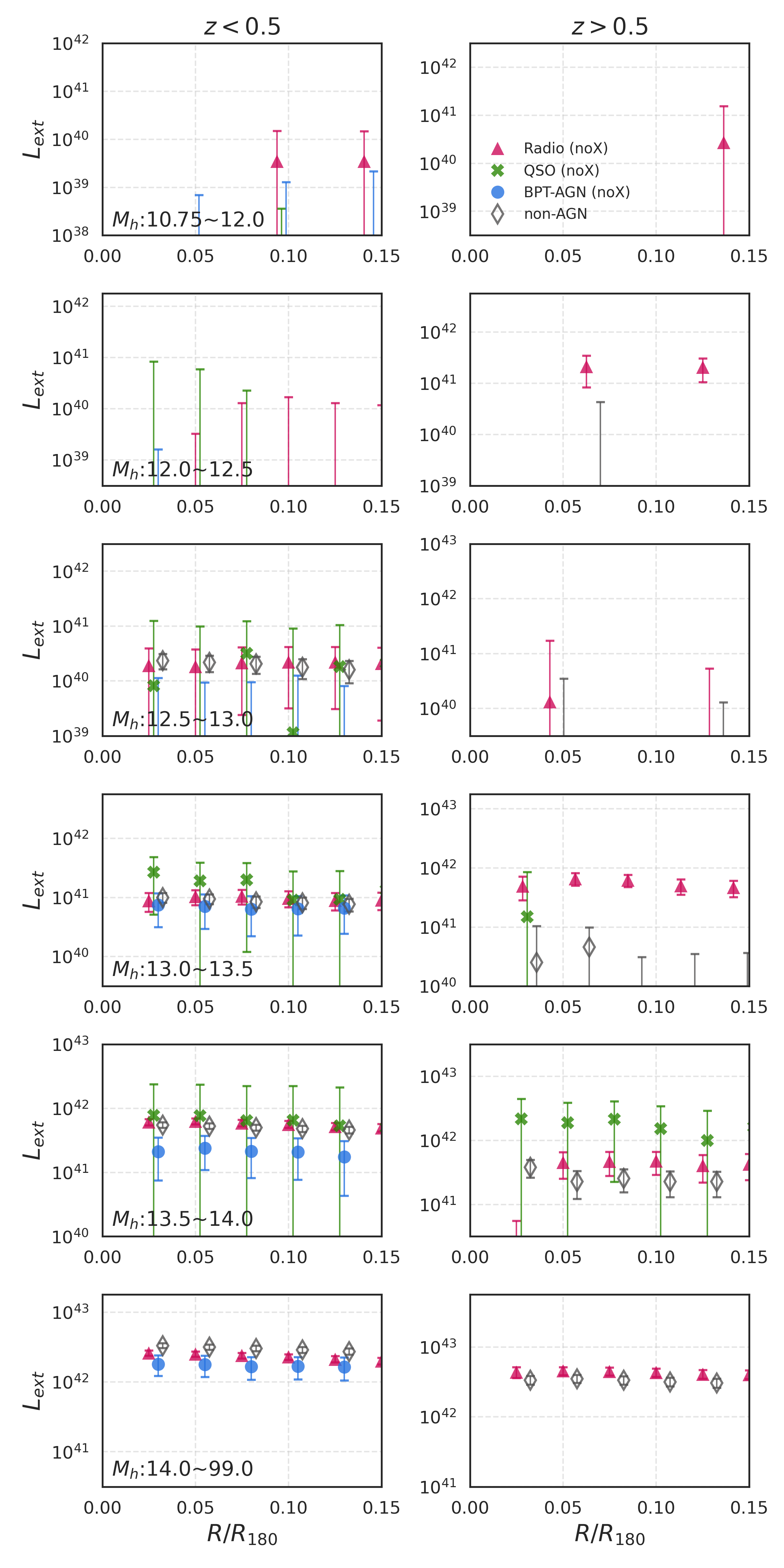}
    \caption{The X-ray emission from the gas component after subtracting the PSF component-accounting for offset effects-is normalized using different central radial ranges. We only show the results for the groups with QSO, Radio, BPT-AGN and non-AGN centrals after removing the X-ray blind-detected sources. The data points have been manually offset along the x-axis for clarity.}
    \label{fig:offsets}
\end{figure}

In Figure~\ref{fig:offsets}, we show that the X-ray emission from the extended gas component after subtracting the PSF component, which accounts for offset effects, is normalized using different central radial ranges. It is evident that the error bars for groups with QSO centrals have once again become large that the gas component becomes indistinguishable. Only in the bin of $z_g < 0.5$ with $M_h \sim 10^{13-13.5}h^{-1}M_\odot$, the robustness remains relatively high, and the extended gas component is also the brightest compared to other samples. For $z_g < 0.5$ with $M_h \sim 10^{13.5-14}h^{-1}M_\odot$, the increase in error is mainly due to the very small sample size (197 groups). The performance in the lower $M_h$ bins further demonstrates that the X-ray contribution from the QSOs is too dominant that the luminosity of the extended gas component is insufficient to be distinguished.

In contrast, the luminosities of the gas components of the other types are much more robust, especially for groups with $M_h > 10^{13}h^{-1}M_\odot$. For $z_g < 0.5$, the difference between the non-X-ray groups with radio and non-AGN centrals is not very significant, while the extended component of the BPT-AGN centrals is slightly lower than that of the other two types. When $z_g > 0.5$, the distinction between non-X-ray groups with radio and non-AGN centrals is prominent only within the $M_h = 10^{13-13.5}h^{-1}M_\odot$ mass range. In a larger mass regime, this difference becomes more ambiguous. The overall level of distant groups with non-AGN centrals is somewhat lower than that of distant non-X-ray groups with radio centrals. This could be attributed to the fact that distant non-AGN centrals encompass all BPT-AGN centrals not identified as X-ray groups, QSO centrals, or radio centrals. As a result, this particular subset might reduce the extended gas luminosity of distant groups with non-AGN centrals.

On the other hand, the composition of the sample without AGN centrals is also quite complex, even for $z_g < 0.5$. Due to the fact that radio centrals are mostly found in massive groups, it does not necessarily mean that massive halos are the only site of radio galaxy formation, although there is certainly some correlation. It is also possible that the RACS catalog is not deep enough, causing some moderately weak radio galaxies to be missed in the selection. Similarly, due to limitations in the spectral signal-to-noise ratio, some weak QSOs might also have been missed.

In summary, after accounting for the offset effect based on the assumption that all observed offsets are caused by systematic observational errors, the gas components of the non-X-ray groups with $z_g < 0.5$ and $M_h \gtrsim 10^{12.5}h^{-1}M_\odot$, or $z_g > 0.5$ and $M_h \gtrsim 10^{13.5}h^{-1}M_\odot$, can still be reliably obtained, except for non-X-ray groups with QSO centrals, where the number of usable samples is very limited. It is important to note again that we are considering the most extreme scenario, where the extended gas luminosity is at its minimum， since we cannot determine whether the offset arises solely from measurement errors, from a genuine displacement between the X-ray peak and the central galaxy, or from a combination of both. The results derived using this method thus only represent a lower limit for the X-ray luminosity for extended gas.

\section{Conclusions}\label{sec:conclusion}

In this study, we investigate the contribution of AGN and extended gas to the total X-ray luminosity, $L_X$, of galaxy groups with different $M_h$ at different redshifts. Although the exposure time of eRASS1 is $\sim 1/10$ that of eFEDS, making it much shallower, eRASS1 covers $\sim 9000$ deg$^2$ of the sky area overlap with the DESI footprint. This allows us to perform X-ray image stacking for $\sim 43$ million galaxy groups with $0 < z_g < 1$ and $M_h \gtrsim 10^{10.75}h^{-1}M_\odot$.

Using the eRASS1 Counterpart Catalog (CTP) and the Rapid ASKAP Continuum Survey (RACS) Catalog, we identified X-ray groups and groups with radio centrals from the entire sample of $\sim 40$ million groups. On the other hand, only $\sim 2.33$ million central galaxies have spectroscopic measurements, with clear emission line measurements or QSO classifications. Therefore, we can identify groups with QSO, BPT-AGN, and non-AGN centrals from this subsample. In the presence of a central AGN, the proportion of BPT-AGN is the highest. The overlap between X-ray groups and groups with QSO centrals is very high, whereas the overlap between groups with radio centrals and other categories is much lower. However, this might be driven by selection effects, which could be related either to eRASS1 pipeline itself or to the multi-wavelength surveys.

We stacked different types of groups based on the AGN classification of their central galaxies. The main results of this paper are summarized as follows:

\begin{enumerate}
    \item By stacking the rest frame $0.5-2.0$ keV X-ray images of the groups without any counterparts in the CPT catalog, we observe a clear X-ray central excess, regardless of the group catagory, redshift, or $M_h$. After all, these subsamples contain central AGNs that are not bright enough in the X-ray to be detected by eRASS1. Even for groups with non-AGN centrals, we can still observe a slight central excess in the low-mass stacked images, except for the lowest $M_h$ bin at $z_g > 0.5$, where the sample size is too small.
    \item We measure the stacked surface brightness profile for different types of groups in different $z_g$ and $M_h$ bins. A general trend is observed whereby extended X-ray emission in the outskirts tends to be more prominent for groups with more massive $M_h$. In contrast, for groups with lower $M_h$, the surface brightness profile are more likely to be represented by the convolved PSF component, such as AGN, alone. This mainly attrributes to the relatively smaller angular size of lower $M_h$ groups compared to the PSF size, our methodology might be biased towards overestimating the contributions of AGN for lower mass groups. On the other hand, at a fixed $M_h$, if the central AGN is too bright in X-ray (e.g., QSOs), it would affect the measurement of the extended X-ray emission in the outskirts. 
    \item When investigating the $L_X$-$M_h$ relations for different types of groups, we find that, overall, $L_X$ increases with increasing $M_h$. Within the same $z_g$ and $M_h$ bin, X-ray groups exhibit the brightest $0.5-2.0$ keV band $L_X$, followed by groups with QSO, radio, BPT-AGN, and non-AGN centrals. Moreover, the lower $M_h$, the larger the differences between these categories. This difference in $L_X$ is primarily driven by the discrepancy in the central AGN component which plays a more dominant role in the low-mass regime.
    \item When excluding the subset containing X-ray detected counterparts from the groups with radio, QSO, and BPT-AGN centrals. Their relative order in $L_X$ remains consistent, and the extended gas component becomes easier to obtain. The gas components of X-ray groups and non-X-ray groups with QSO centrals are the brightest, significantly higher than those of the other categories. In addition, the extended gas component for those other categories have a very similar $L_X-M_h$ scaling relation. 
    \item If we account for the optical and X-ray positional offset effects, assuming that these offsets are entirely due to systematic errors, the gas component of non-X-ray groups with QSO centrals is significantly reduced. Given the extremely small sample size for this category (due to the low proportion of QSO and the limited subsample of $\sim 2.33$ million groups), the error bars are extremely large.
\end{enumerate}

Currently, we cannot confidently assert that, within the same $z_g$ and $M_h$ bin, the higher total $L_X$ (or X-ray luminosity of the central AGN) necessarily corresponds to an X-ray brighter gas component. This correlation vanishes only if all positional offset effects are ascribed to systematic errors, representing the most adverse condition for affirming a correlation between the overall and extended gas X-ray luminosity. Once more reliable evidence emerges showing that the positional offset effect is largely a result of genuine offsets caused by AGN feedback, or if more DESI spectroscopic measurements and deeper data releases from eRASS become available in the future, the correlation between the two could potentially be re-established. 

Regardless, it is evident that the integration of eROSITA's all-sky X-ray observations with a more comprehensive multi-wavelength galaxy group sample offers substantial potential for precisely quantifying the contributions of different components to the X-ray emission, particularly for groups in the low-mass regime.

\section*{Acknowledgements}
This work is supported by the National Key R\&D Program of China (2023YFA1607800, 2023YFA1607804), “the Fundamental Research Funds for the Central Universities”, 111 project No. B20019, and Shanghai Natural Science Foundation, grant No.19ZR1466800. We acknowledge the science research grants from the China Manned Space Project with Nos. CMS-CSST-2021-A02 \& CMS-CSST-2025-A04, and Yangyang Development Fund.
The computations in this paper were run on the Gravity Supercomputer at Shanghai Jiao Tong University. This project is also supported in part by Office of Science and Technology, Shanghai Municipal Government (grant Nos. 24DX1400100，ZJ2023-ZD-001), and China Postdoctoral Science Foundation (2023M742286).

This research used data obtained with the Dark Energy Spectroscopic Instrument (DESI). DESI construction and operations is managed by the Lawrence Berkeley National Laboratory. This material is based upon work supported by the U.S. Department of Energy, Office of Science, Office of High-Energy Physics, under Contract No. DEAC0205CH11231, and by the National Energy Research Scientific Computing Center, a DOE Office of Science User Facility under the same contract. Additional support for DESI was provided by the U.S. National Science Foundation (NSF), Division of Astronomical Sciences under Contract No. AST-0950945 to the NSF’s National Optical-Infrared Astronomy Research Laboratory; the Science and Technology Facilities Council of the United Kingdom; the Gordon and Betty Moore Foundation; the HeisingSimons Foundation; the French Alternative Energies and Atomic Energy Commission (CEA); the National Council of Science and Technology of Mexico (CONACYT); the Ministry of Science and Innovation of Spain (MICINN), and by the DESI Member Institutions: www.desi.lbl.gov/collaborating-institutions.

The DESI Legacy Imaging Surveys consist of three individual and complementary projects: the Dark Energy Camera Legacy Survey (DECaLS), the Beijing-Arizona Sky Survey (BASS), and the Mayall z-band Legacy Survey (MzLS). DECaLS, BASS and MzLS together include data obtained, respectively, at the Blanco telescope, Cerro Tololo Inter-American Observatory, NSF’s NOIRLab; the Bok telescope, Steward Observatory, University of Arizona; and the Mayall telescope, Kitt Peak National Observatory, NOIRLab. NOIRLab is operated by the Association of Universities for Research in Astronomy (AURA) under a cooperative agreement with the National Science Foundation. Pipeline processing and analyses of the data were supported by NOIRLab and the Lawrence Berkeley National Laboratory (LBNL). Legacy Surveys also uses data products from the Near-Earth Object Wide-field Infrared Survey Explorer (NEOWISE), a project of the Jet Propulsion Laboratory/California Institute of Technology, funded by the National Aeronautics and Space Administration. Legacy Surveys was supported by: the Director, Office of Science, Office of High Energy Physics of the U.S. Department of Energy; the National Energy Research Scientific Computing Center, a DOE Office of Science User Facility; the U.S. National Science Foundation, Division of Astronomical Sciences; the National Astronomical Observatories of China, the Chinese Academy of Sciences and the Chinese National Natural Science Foundation. LBNL is managed by the Regents of the University of California under contract to the U.S. Department of Energy. The complete acknowledgments can be found at https://www.legacysurvey.org/acknowledgment/.

This work is also based on data from eROSITA, the soft X-ray instrument aboard SRG, a joint Russian-German science mission supported by the Russian Space Agency (Roskosmos), in the interests of the Russian Academy of Sciences represented by its Space Research Institute (IKI), and the Deutsches Zentrum f\"{u}r Luftund Raumfahrt (DLR). The SRG spacecraft was built by Lavochkin Association (NPOL) and its subcontractors, and is operated by NPOL with support from the Max Planck Institute for Extraterrestrial Physics (MPE). The development and construction of the eROSITA X-ray instrument was led by MPE, with contributions from the Dr. Karl Remeis Observatory Bamberg \& ECAP (FAU Erlangen-Nuernberg), the University of Hamburg Observatory, the Leibniz Institute for Astrophysics Potsdam (AIP), and the Institute for Astronomy and Astrophysics of the University of T\"{u}bingen, with the support of DLR and the Max Planck Society. The Argelander Institute for Astronomy of the University of Bonn and the Ludwig Maximilians Universit\"{a}t Munich also participated in the science preparation for eROSITA. The eROSITA data shown here were processed using the eSASS software system developed by the German eROSITA consortium.

\begin{appendix}
\section{X-Ray Image Stacking} \label{sec:im_stacking}

For each group, we retrieve all events located within a distance of $2\theta_{180}$ from the X-ray center. We corrected the observed energy ($E_{\rm obs}$) of all photons to the rest frame energy: $E_{\rm rf} = (1+z_g)E_{\rm obs}$. We rescaled the images of the group in different $M_h$ bins to a uniform size. To provide the projected profile of the luminosity of the stacked surface in erg / s / kpc$^2$, each event is weighted by the following.
\begin{equation}
    W = g\left(n_{\rm H}, E_{\rm obs}\right) \cdot \frac{1.602 \times 10^{-9} \cdot E_{\rm{rf}}}{{\rm ARF} \left({E}_{\rm{obs}} \right)} \cdot \frac{4\pi d_{\rm{L}}^{2}}{t_{{\rm exp0}}} \,\left(\rm{erg/s}\right).
\end{equation}
Here, $g\left(n_{\rm H}, E_{\rm obs}\right)$ is a function that corrects for absorption effects, with $n_{\rm H}$ taken from \citet{Kalberla..2005}. For events with $E_{\rm obs} \lesssim 0.25$ keV, $g\left(n_{\rm H}, E_{\rm obs}\right)$ is extremely sensitive to $n_{\rm H}$, which can easily lead to catastrophic value and reduce the signal-to-noise ratio of stacking results. Therefore, we focus only on events with a rest frame energy of $0.5 {\rm \,keV} \le {E}_{\rm{rf}} \le 2.0 \rm{\,keV}$. The ${\rm ARF} \left({E}_{\rm{obs}} \right)$ is the effective area of the telescope as a function of energy. In each tile, ${\rm ARF} \left({E}_{\rm{obs}} \right)$ varies, particularly at the observed energies $E_{\rm obs} \lesssim 0.25$ keV. To account for potential variations due to the differences in the ARF, we use the ARF corresponding to the central RA and Dec of each tile for groups within the same tile, rather than adopting a survey-averaged ARF. Note that the ARF value used here is given by dividing the ``SPECRESP" by the correction factor ``CORRCOMB". The $d_{\rm{L}}$ is the luminosity distance for each group and $t_{\rm exp0}$ is the exposure time that has not been corrected by vignetting, respectively. The value $1.602 \times 10^{-9}$ is the conversion factor from keV to erg.

Consequently, each pixel in the stacked image also needs to be assigned a weight to represent the area size corresponding to that pixel. This is necessary because during the stacking process, scaling is involved, and the area represented by the same pixel might vary. For the pixel in the $i$-th row and the $j$-th column, the weight value, $A_{\rm i,j}$, is given by:
\begin{equation}
    A_{\rm i,j} = \sum_{k=1}^{N} \left[\frac{R_{180,k}\left(\rm{kpc}\right)}{R_{\rm{pix}}\left(\rm{pix}\right)}\right]^{2}\cdot \delta_{k}\left(i,j\right),
\end{equation}
$\delta_{k}\left(i,j\right)$ indicates whether the $k$-th group has been masked in this pixel, and $R_{\rm{pix}}$ is the fixed pixel size of the stacked image. 

The X-ray background for each group is determined using an annulus in the range of $1.5R_{180} < R < 2R_{180}$. The background level is determined by summing the event weights within this annulus and dividing by the sum of $A_{\rm i,j}$ in the same region. 

\subsection{Masking of Contaminants}
Since stacking might introduce contamination from foreground or background X-ray sources in each individual image, we applied contaminant masking based on the CTP catalog matching results (see details in Section~\ref{sec:xmaps}). Specifically, all blind-detected X-ray sources that fall within $2\theta_{180}$ of the center of the group but are not associated with the central of that group are masked. 

As suggested by \citet{Comparat..2023}, the masking radius, $\theta_{\rm src}$, for each blind-detected source is related to its photon counts (see Figure A.1 of \citet{Comparat..2023}). To be conservative, we adopt a masking radius of $1.4\theta_{\rm src}$ for point sources and $1.2\theta_{\rm src}$ for extended sources. The minimum of the masking radius is $\sim 30"$. This masking procedure leads to variations in the weight values $A_{i,j}$ assigned to different pixels in the stacked image.

\subsection{X-ray Profile and Luminosity}

The X-ray profile, $S_{X}\left(R\right)$, as a function of $R$, is then obtained by summing the event weights within $R-0.5\Delta R$ to $R+0.5\Delta R$ and dividing by the sum of $A_{\rm i,j}$ in this range, followed by subtracting the background level. For each stacked image, the number of radius bins is equal to twice the median $\theta_{180}$ (in arcsec) of the groups being stacked (at most 40 bins), as Figure~\ref{fig:xprofile} shows.

The total X-ray luminosity, $L_X$, for the stacked groups is obtained by summing all events corrected for the loss in $A_{\rm i,j}$ due to masking within $R_{180}$ and then dividing by the number of groups that were stacked. However, due to the higher uncertainty in X-ray brightness at the edges of the groups, calculating $L_X$ by summing all the corrected events within $R_{180}$ might introduce significant uncertainty.

\begin{figure*}
    \centering
    \includegraphics[width=0.475\linewidth]{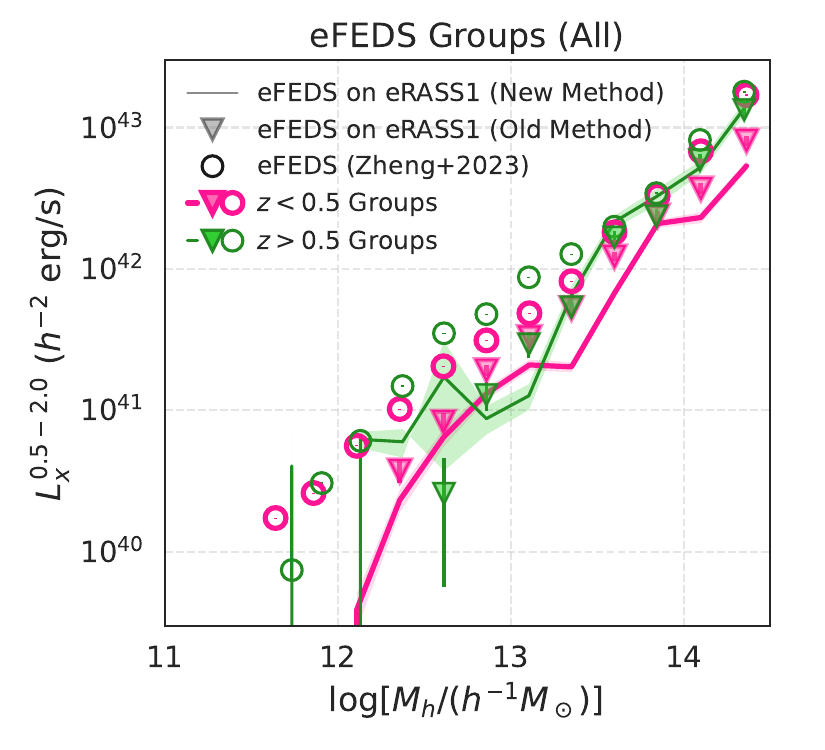}
    \includegraphics[width=0.475\linewidth]{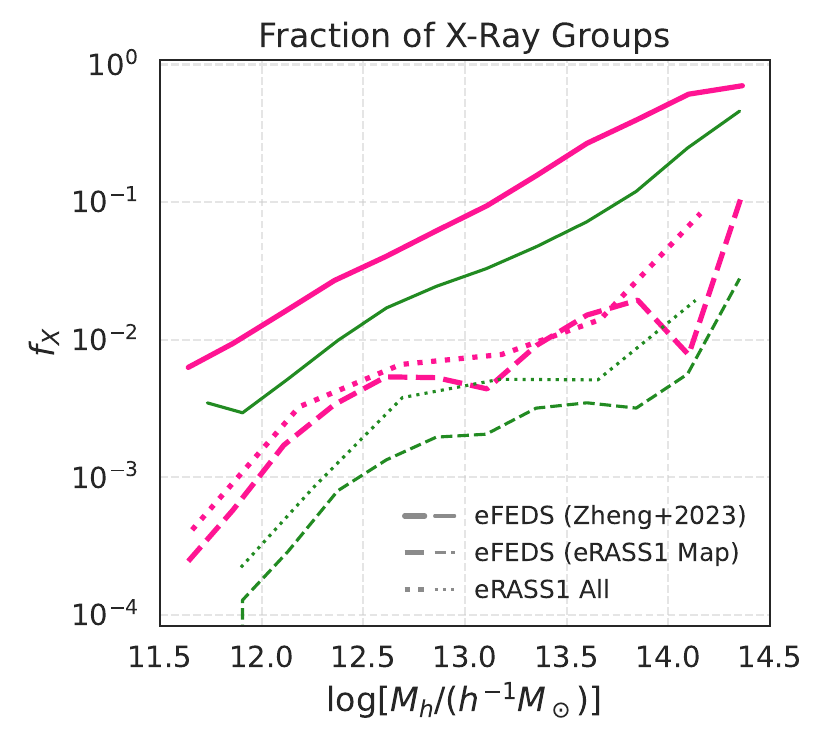}
    \includegraphics[width=0.475\linewidth]{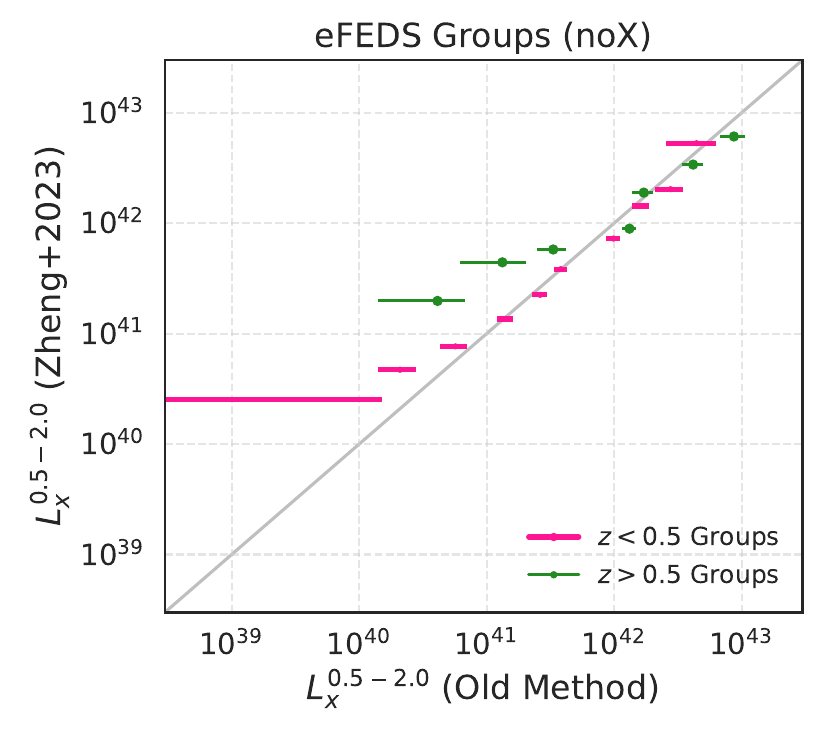}
    \includegraphics[width=0.475\linewidth]{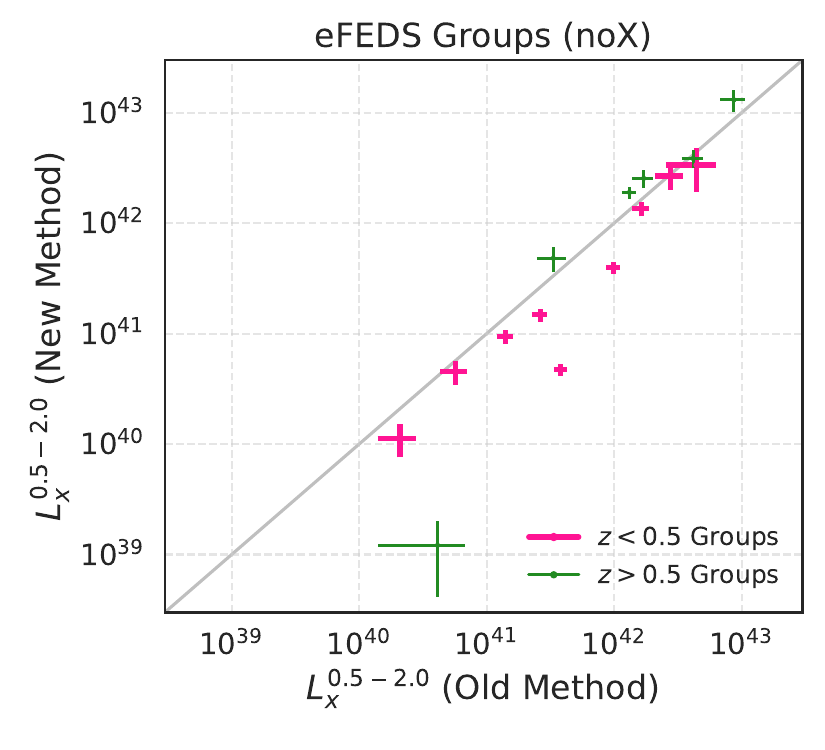}
    \caption{Upper Left: The $L_{\rm X}-M_h$ relations for the DESI groups overlaid on eFEDS footprint are shown using the new method described in Appendix~\ref{sec:im_stacking} (solid lines with shaded regions) and the previous method of \citet{Zheng..2023} (triangles with error bars), both derived from the eRASS1 map. The results of eFEDS groups derived from the eFEDS map are shown in circles with error bars. Upper Right: The fraction of X-ray groups at different $M_h$ bins for groups overlaid on eFEDS footprint as given by \citet{Zheng..2023} (solid lines) and eRASS1 map in this work (dashed lines). We also show the fraction of X-ray groups in the full sample (dotted lines). Lower Left: A comparison of the $L_{\rm X}$ values for the eFEDS groups not detected as X-ray sources in the eFEDS or eRASS1 maps, as obtained in this study and by \citet{Zheng..2023}. Both values were derived using the old method. Lower Right: Similar to the lower left panel, but comparing the results derived using the old and new methods based on the eRASS1 map.}
    \label{fig:efeds}
\end{figure*}

\subsection{Error Estimate}

As suggested by \citet{Zhang..2024a}, the uncertainties in $S_X$ and $L_X$ consist of two parts: the Poisson fluctuation of the data and the uncertainty from the stacking process. The Poisson fluctuation can be obtained from the signal-to-noise ratio, $S/N$, using
\begin{equation}
    S/N = \frac{N_{\rm src}}{\sqrt{N_{\rm tot}+N_{\rm bkg}}},
\end{equation}
where $N_{\rm bkg}$ is the background counts scaled to the aperture of radius $R_{X}$, and $N_{\rm src}$ is the net photon counts within the aperture radius.

The uncertainty of the stacking process can be computed using the Jackknife resampling method \citep{Andrae..2010, McIntosh..2016} to assess the variation of the stacked samples. For a galaxy sample containing $N$ galaxies, we randomly selected $90 \%$ of them to stack 50 times and retrieved the standard deviation of these 50 stacking results.

In general, the uncertainty from the stacking process is slightly smaller than the Poisson fluctuation. This is primarily caused by low number counts for each source which leads to a high Poisson noise. If the sample size for stacking or the exposure time is sufficiently long, leading to a higher photon count, the uncertainty from the stacking process would become more dominant. In summary, the total uncertainty is the square root of the sum of the squares of these two parts.

\subsection{Differences Between the Method Used in This Study and \citet{Zheng..2023}}

The stacking method in this study differs significantly from the mean background subtraction algorithm introduced in \citet{Zheng..2023}. In short, the old method calculates the count rates in $0.2-2.3$ keV and then uses an energy conversion factor (ECF) to convert them to X-ray flux. This factor is derived by assuming the X-ray spectrum of a group and, after applying eROSITA’s ARF and RMF, obtaining a theoretical count rate. The ECF is then the ratio between the observed count rate and the theoretical count rate. In \citet{Zheng..2023}, the ECF is based on the assumption of a power law with a slope of 2.0.

In the upper left panel of Figure~\ref{fig:efeds}, we compare the $L_{\rm X}-M_h$ relations for the groups overlaid on eFEDS footprint obtained in this study and by \citet{Zheng..2023}. It is evident that the relation of \citet{Zheng..2023} is overall approximately $\sim 0.5$ dex higher than the results of this study, especially for the groups at $z_g < 0.5$. This appears to suggest a possible inconsistency between the two methods. However, if we apply the old method to the groups overlaid on eFEDS using the eRASS1 map, it becomes clear that the results produced by the same method are still lower than those of previous. Meanwhile, the results of the old and new methods are generally quite consistent with each other, except for a very small discrepancy that might be attributed to the different assumptions underlying the ECF. It is worth noting that, as shown in Figure~\ref{fig:LX3}, the enhancement of overall $L_X$ due to X-ray groups could be significant across all $M_h$ bins, especially in the low-mass regime. 

In the upper right panel of Figure~\ref{fig:efeds}, we plot the X-ray group fraction for the groups overlaid on eFEDS footprint as reported by \citet{Zheng..2023} and by this study, as well as the fraction of X-ray groups in full sample. Although the fraction of X-ray groups tends to be very low for groups with $M_h \lesssim 10^{13}h^{-1}M_\odot$, the results of \citet{Zheng..2023} are still in general $\sim 1$ dex higher than those from this study. There are two main reasons for this. First, the average vignetted exposure time of eFEDS is $\sim 1300$ seconds, while the vignetted eRASS1 exposure time in the same area is $\sim 100$ seconds. In fact, across most of the eRASS1 sky, except for a small region near the South Ecliptic Pole, the exposure time remains at this very low level, leading to a significantly reduced number of blind-detected X-ray sources in eRASS1. Second, the CTP catalog adopts a more stringent criterion for associating X-ray sources with galaxy groups, which also contributes to the lower detected counts.

If we remove all blind-detected X-ray groups in eFEDS, regardless of whether they were detected in the eFEDS or eRASS1 and then bin the remaining groups by different $M_h$, we can repeat the above test by comparing the results from: (1) the old method applied to eFEDS, (2) the old method applied to eRASS1, and (3) the new method applied to eRASS1. The lower panels of Figure~\ref{fig:efeds} indicate that the outcomes between the two initial cases are largely consistent, except at the faintest end. The notable variations observed in the lower $L_X$ region are due to the greater uncertainty in the background level estimated from the eRASS1. This is due to a considerably shorter exposure time that results in reduced signal, which accounts for the larger error bars in the eRASS1 results. However, when comparing the latter two cases, it is also clear that, except for a few isolated bins, the results are highly consistent.

This fundamentally illustrates the self-consistency of both methods. The variations in the $L_X-M_h$ relation are more apt to be influenced by the criteria employed to allocate X-ray sources, which can considerably affect the $L_X-M_h$ relation, particularly at the lower mass spectrum.

\begin{figure*}
    \centering
    \includegraphics[width=0.49\linewidth]{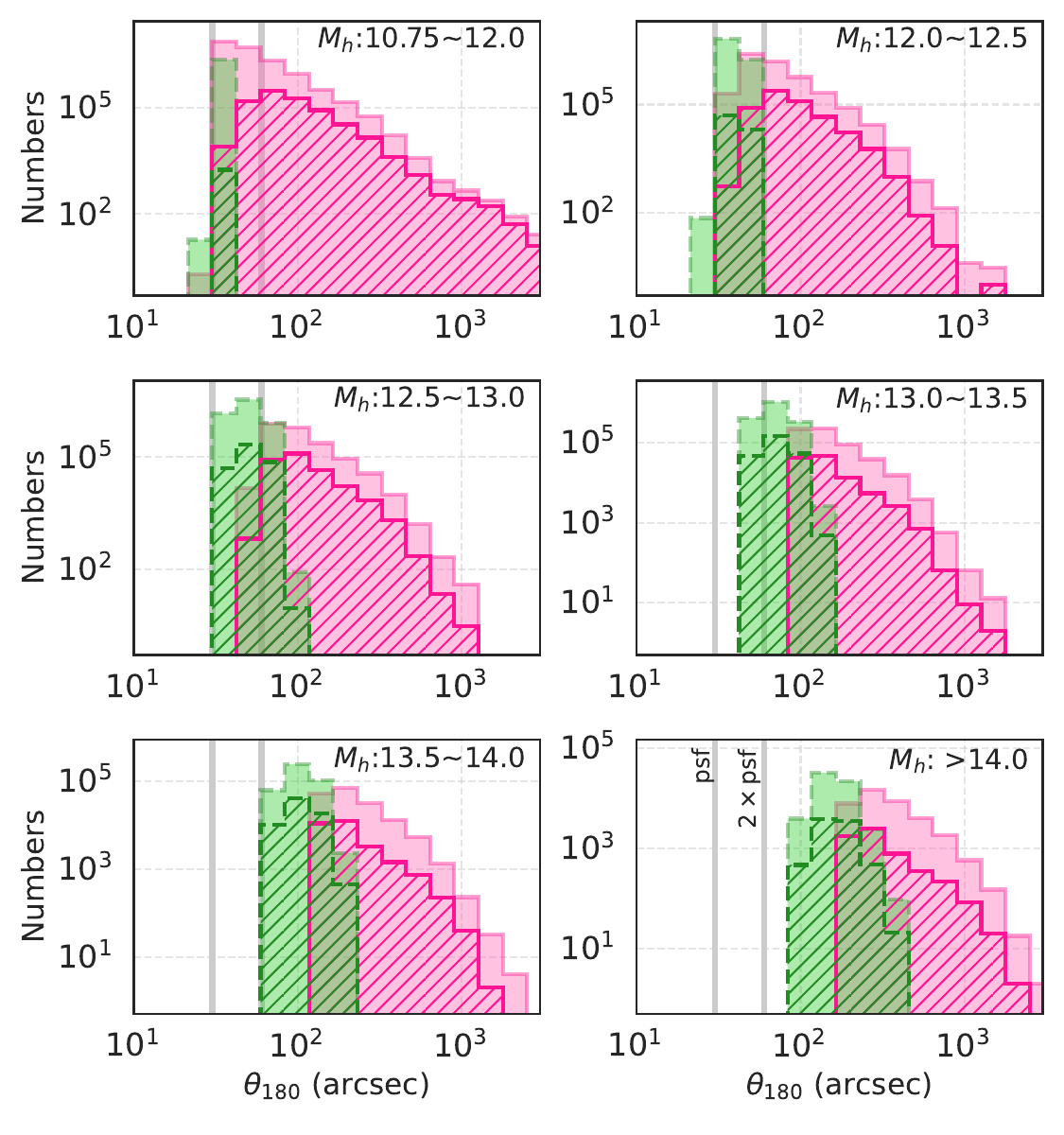}
    \includegraphics[width=0.49\linewidth]{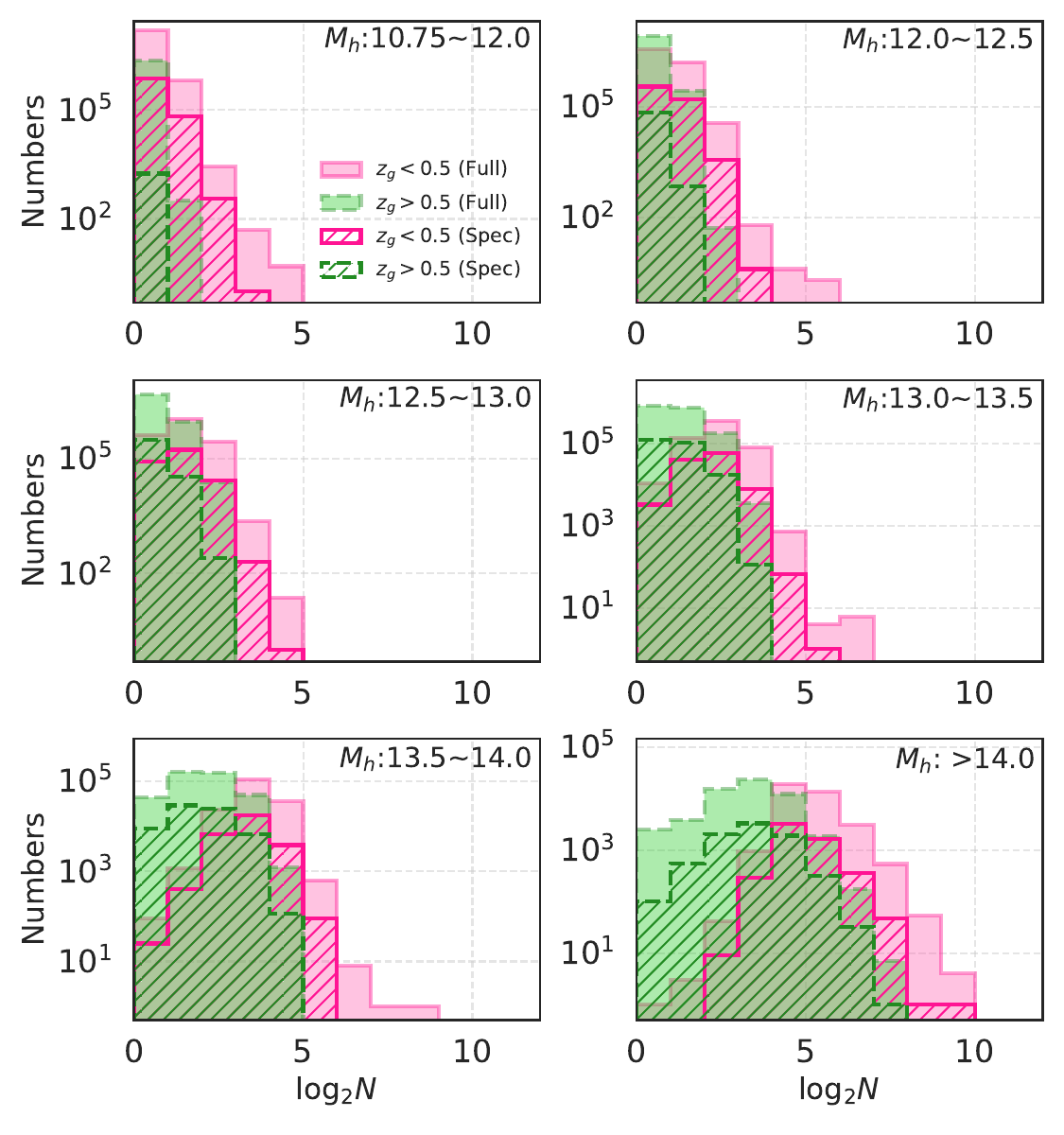}
    \caption{The $\theta_{180}$ (left) and richness (right) distributions for DESI groups at different redshift and $M_h$ bin. The green and pink histograms represent the distributions for groups at $z_g < 0.5$ and $z_g > 0.5$, respectively, while the filled and hatched styles correspond to the full and spectroscopic samples. The grey vertical lines in the left panel mark the PSF size ($30''$, containing $\sim 72\%$ of the PSF) and its twofold (containing $\sim 88\%$ of the PSF) for point sources in eROSITA pipeline.}
    \label{fig:theta180}
\end{figure*}

\section{Contribution and Contamination from Point Sources}\label{sec:psf}

To determine the contributions of different components to the X-ray emission, it is necessary to model both the AGN and the X-ray binaries. Given that both AGNs and X-ray binaries are intrinsically point-like sources, we model these two components using the eROSITA point spread function (PSF) in this section. On the basis of the predicted luminosities of the X-ray binaries and AGN, we normalize the PSF to obtain their X-ray extended emission.

\subsection{Central AGN and X-ray Binary}
We used eROSITA PSF taken from \citet{Sanders..2025} to model the central AGN and X-ray binary. Note that the X-ray emission from X-ray binaries in galaxies is contributed by their stellar component and should not be treated as a point source. However, the typical half-light radius for massive galaxies is $r_{50} \simeq 10$ kpc \citep{Shen..2003}, which is negligible compared to $R_{180}$ for most groups used in this work (see the bottom row of Figure~\ref{fig:xstack}). Therefore, for simplicity, we model both the AGN and X-ray binaries from central galaxies using the eROSITA PSF.

For each subsample, we convolve the PSF profile with the scaling factor given by the ratio of the $\theta_{180}$ (in pixels) of each PSF’s corresponding group and the fixed pixel size, $R_{\rm{pix}}$, of the stacked image. We can naively assume that all central AGN and X-ray binary in each group of a subsample have the same X-ray luminosity. In other words, during the convolution process, each member contributes equally to the PSF. However, many of our subsamples contain X-ray-detected groups whose X-ray luminosities are too high that their point source components can significantly affect the overall convolved PSF. Therefore, we use two PSF models here: one assumes that the luminosity of the central point source is the same (unweighted) for all groups in the same sub-sample, while the other is weighted by the individual luminosity of each group. The individual X-ray luminosity of each group is not calculated within $\theta_{180}$, but rather within the PSF half-energy width of a radius $30''$.

To accurately resolve the extended X-ray emission, the angular size of the group must be taken into account. If $\theta_{180}$ is too small, the extended component will be unsuitably resolved. In the left panel of Figure~\ref{fig:theta180}, we show the $\theta_{180}$ distributions for the DESI groups at different redshift and $M_h$ bins. As a reference, we also mark the size of the eROSITA PSF as $30''$ \citep[containing $\sim 72\%$ of the PSF, ][]{Merloni..2024, Sanders..2025}. It is evident that $\theta_{180}$ for the majority of groups exceed the size of the PSF, with most of the $M_h \gtrsim 10^{12.5}h^{-1}M_\odot$ groups reaching twice the size of the PSF or more. In contrast, the groups with $z_g > 0.5$ and $M_h \lesssim 10^{12}h^{-1}M_\odot$ exhibit the smallest $\theta_{180}$. However, as shown in Figure~\ref{fig:xprofile}, this subset of groups will not significantly affect our results due to their very limited sample size within this bin.

Here, it's crucial to acknowledge that we do not distinguish the origin of the central point source contribution, whether it stems from AGN or X-ray binary. In most subsamples featuring AGN components, the luminosity from the PSF component notably exceeds the contribution from X-ray binaries. The extent of the X-ray emission was evaluated by scaling the mean PSF profile to match the observed central surface brightness, thereby deriving a maximal PSF-like surface brightness profile. Should an excess of emission, surpassing the maximal PSF-like profile, be detected at larger radii, it is indicative that the X-ray emission originates from a source extending beyond the confines of the PSF, such as the extended warm/hot gas.

\begin{figure}
    \centering
    \includegraphics[width=1.\hsize]{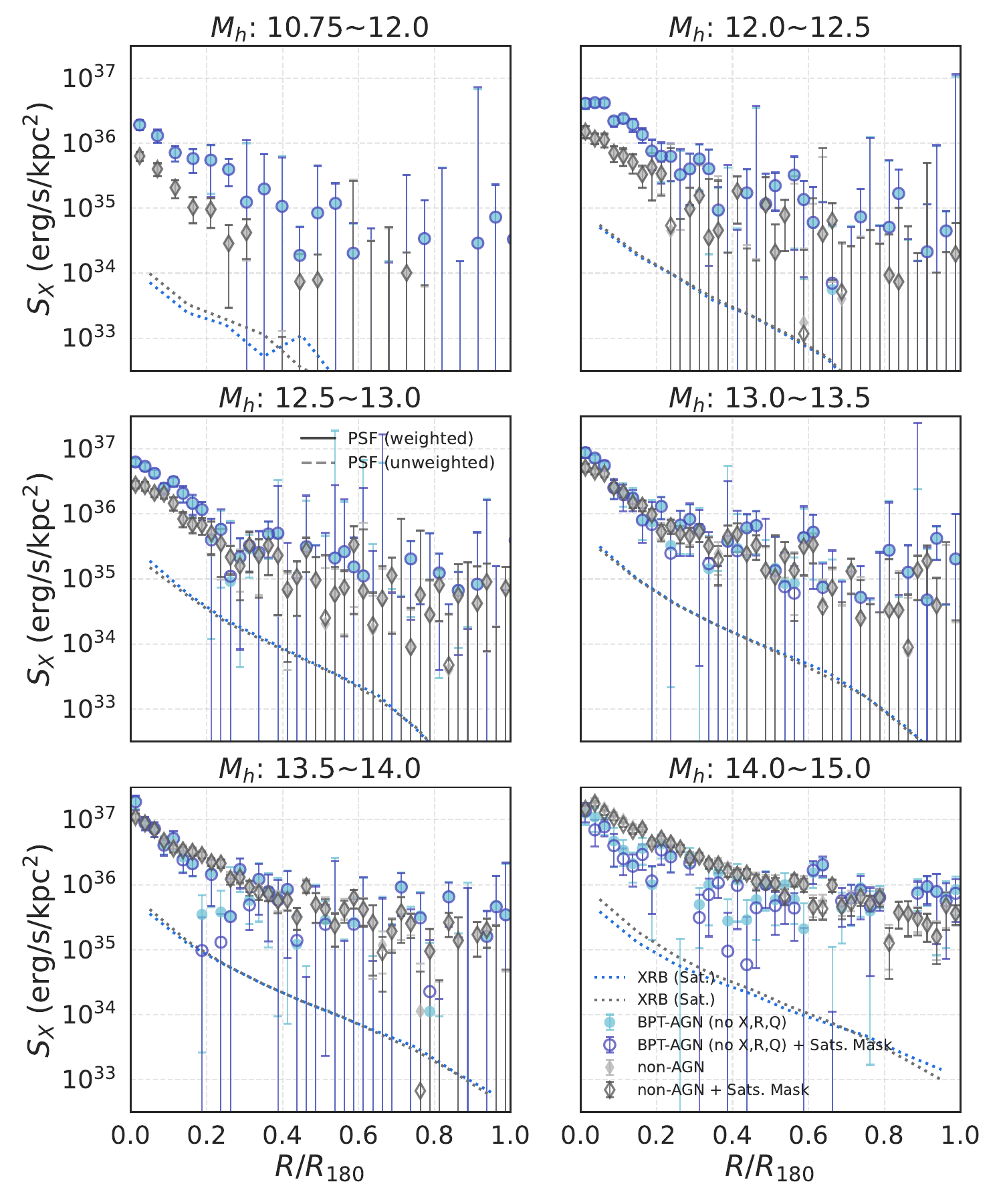}
    \caption{X-ray surface brightness profiles for groups with non-AGN central (grey diamond) and BPT-AGN central (blue circle) after removing the overlaps with X-ray groups, groups with QSO centrals and radio centrals. The filled symbols represent the profile obtained from satellite unmasked images, while the open symbols represent the stacked images after masking all potential contaminated satellites. The dotted lines are the contribution from X-ray binaries in satellites.}
    \label{fig:xrbprof}
\end{figure}

\begin{figure}
    \centering
    \includegraphics[width=1.\hsize]{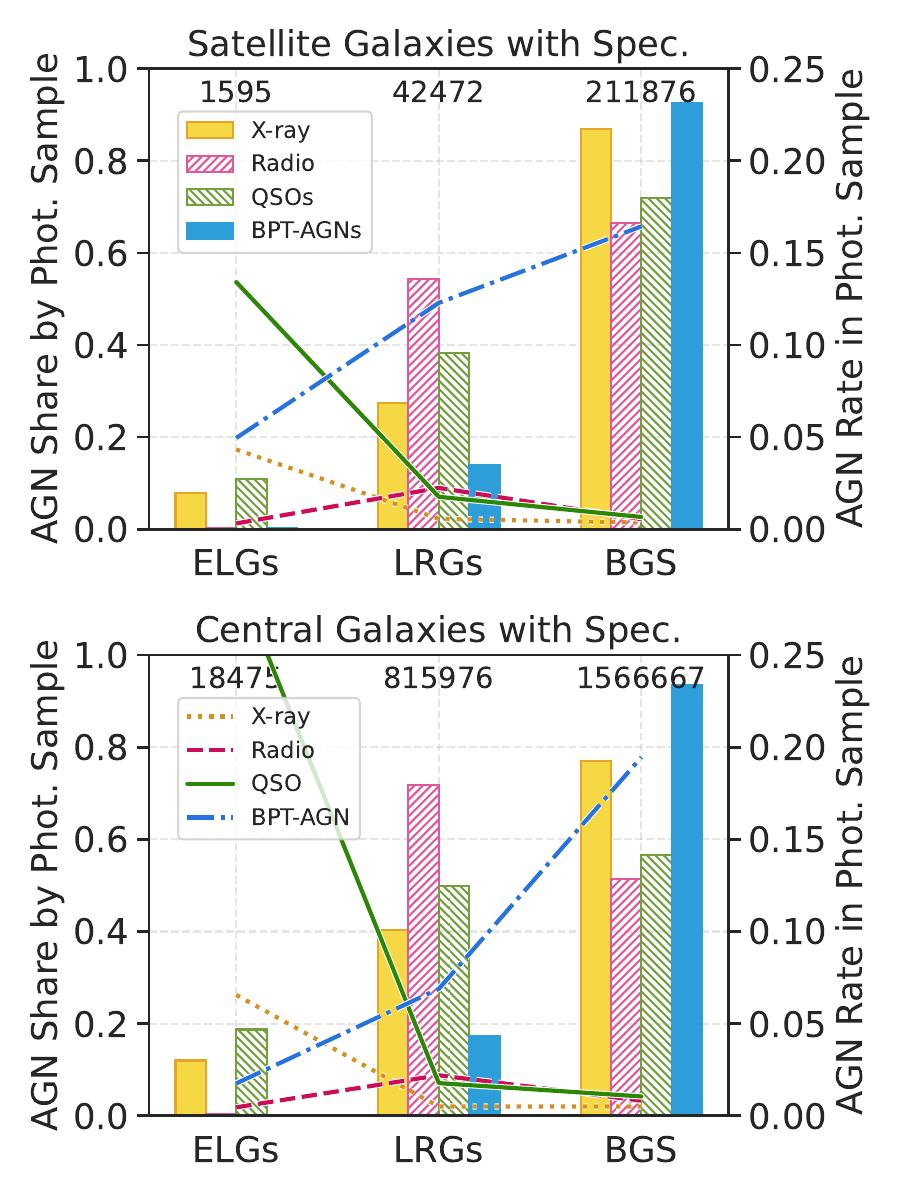}
    \caption{Distribution of AGNs across different photometric subsamples (ELGs, LRGs, and BGS) for galaxies with spectroscopic observations. The histograms show the fractional contribution of each photometric subsample to the total X-ray groups (yellow filled), Radio (red hatched), QSOs (green hatched), and BPT-AGNs (blue filled), while the solid lines display the AGN fraction within each subsample. Note that due to overlaps between photometric subsamples, the summed contributions from different subsamples for a given AGN type might exceed 1. The number of different types of photometric subsamples is labeled at the top of each panel.}
    \label{fig:shares}
\end{figure}

\subsection{Contribution and Contamination from Satellite Galaxies}\label{sec:satcontri}

\subsubsection{Satellite AGNs}

When stacking groups, the X-ray emission from X-ray binaries and AGNs for satellite galaxies might be confused with the emission from the extended component. In the right panel of Figure~\ref{fig:theta180}, we plot the richness distribution of groups at different $M_h$. For $M_h \gtrsim 10^{12.5}h^{-1}M_\odot$, most of the groups contain satellite galaxies, particularly those with higher $M_h$. This implies that low $M_h$ groups are less likely to be contaminated by satellites. However, for groups with higher richness, satellite contamination must be taken into account.

As mentioned in Section~\ref{sec:spec}, only $\sim 190$ k of $\sim 2$ million satellite galaxies have spectroscopic measurements. As a result, it is more challenging to identify AGNs residing in satellite galaxies. Among all AGNs, those that have X-ray counterparts are naturally the brightest. However, X-ray point sources that are associated with satellite galaxies have already been masked as contaminants during the stacking process. Consequently, the potential presence of X-ray AGNs in satellite galaxies is not a significant concern. Secondly, since the sample of radio galaxies is complete, this category in the satellite can be easily identified. 

The other two categories that are relatively difficult to identify in satellite galaxies are QSOs and BPT-AGNs, as both heavily rely on DESI spectroscopic measurements. Based on the $\sim 190$k spectroscopically confirmed satellite galaxies, we find that QSOs account for only $\sim 0.7\%$, while BPT-AGN make up $\sim 14.5\%$. After removing the portions that overlap with other categories, the fractions of these two classes drop to $\sim 0.5\%$ and $\sim 13.7\%$, respectively. The abundance of BPT-AGN appears to be quite high, suggesting that the masking of BPT-AGN could be quite problematic. In Figure~\ref{fig:xrbprof}, we plot the X-ray surface brightness profile for non-AGN centrals and BPT-AGN centrals after removing the overlaps with X-ray groups, groups with QSO centrals and radio centrals. Clearly, we see the remaining BPT-AGN centrals display markedly weaker X-ray luminosity. Therefore, we are not necessary to consider the remaining influence of BPT-AGNs.

On the other hand, although the fraction of QSOs is very low, their intrinsic X-ray luminosities are relatively high, and they might still exert a non-negligible impact on the outskirts of groups with intrinsically weak AGN emission.

An extremely straightforward approach would be to mask out all satellite galaxies. However, the consequence of this method is that for groups with even slightly larger $M_h$, an excessive amount of area would be masked out. This would result in the weight value, $A_{\rm i,j}$, for the pixel in the $i$-th row and $j$-th column becoming too small, leading to a significant reduction in the signal-to-noise ratio. 

A more balanced approach is to leverage samples selected for spectroscopic observation based on photometric criteria, including the Bright Galaxy Survey \citep[\texttt{BGS},][]{Ruiz-Macias..2020}, the Luminous Red Galaxies \citep[\texttt{LRG},][]{ZhouRP..2020, ZhouRP..2023}, and the Emission Line Galaxies \citep[\texttt{ELG},][]{Raichoor..2020, Raichoor..2023}. By analyzing both the distribution of AGN types across different photometric subsamples and the probability that each photometric class hosts a given type of AGN, we can make a more comprehensive assessment of their relative contributions. 

In Figure~\ref{fig:shares}, we show the fractional contribution of each photometric subsample to the X-ray counterparts, radio galaxies, QSOs, and BPT-AGNs, respectively. Note that only targets with available spectroscopic observations are included in this analysis. It is evident that, with the exception of radio centrals, the BGS subsample accounts for the largest share across all categories, followed by LRG and then ELG, which contributes the least. However, this predominance mainly reflects the larger sample size of BGS, rather than a higher efficiency in uncovering potential AGN targets. If we were to mask out all the BGS based on their large contributions, the results would be close to masking all satellites, which results in a significantly reduced $A_{i,j}$ across the stacked image. 

The motivation behind employing photometric subsamples to mask satellites was to maximize masking efficiency while minimizing the number of required targets. When examining the occurrence rates of different AGN types, particularly QSOs, within each photometric subsample, it becomes evident that the ELG sample exhibits a notably high QSO detection rate, even though it accounts for only $10 \sim 20\%$ of the total in absolute terms. By comparison, LRG exhibits a relatively modest detection rate, but contributes a considerable fraction to the total number of detections.

Therefore, a conservative approach is to mask out all known AGNs (X-ray sources, radio galaxies, QSOs, and BPT-AGNs), as well as ELG and LRG satellites. After all, ELG and LRG satellites constitute only a small fraction of the overall satellite galaxy population. For QSO satellites, although ELGs and LRGs together contribute less than half, this still implies that slightly more than half of the QSO satellites remain unmasked, which introduces a potential risk of overestimating the extended X-ray luminosity. However, our tests show that including or excluding ELG and LRG satellites in the masking process does not lead to a significant change in the estimated extended X-ray luminosity.

\subsubsection{Satellite X-ray Binaries}
We also plot the radial profile of the X-ray binaries in satellites for groups with non-AGN and BPT-AGN centrals after removing the overlaps with X-ray groups, groups with QSO centrals and radio centrals in Figure~\ref{fig:xrbprof}. It is clear that, compared to the overall surface brightness profile, the contribution of X-ray binaries in satellite galaxies is at least $> 1$ dex lower. The presence of X-ray binaries in satellite galaxies has a minimal influence on our estimation of extended X-ray emission and does not alter the overall conclusions.


\end{appendix}

\end{CJK*}
\end{document}